\documentclass[11pt]{article}
\usepackage{epsf}\usepackage{amssymb} \usepackage{cite}
\usepackage{amsthm}
\usepackage{amscd}
\usepackage{amsfonts}
\usepackage{amsmath}

\newcommand{\be}{\begin{equation}}
\newcommand{\ee}{\end{equation}}
\newcommand{\bea}{\begin{eqnarray}}
\newcommand{\eea}{\end{eqnarray}}

\usepackage{hyperref}
\def\stackreb#1#2{\ \mathrel{\mathop{#1}\limits_{#2}}}

\begin{document}
\thispagestyle{empty}
\def\thefootnote{\fnsymbol{footnote}}
\begin{center}\Large \bf
A parafermionic hypergeometric function\\  and supersymmetric $6j$-symbols \\
\end{center}

\vskip 0.2cm

\begin{center}
Elena Apresyan$^{1}$\footnote{elena-apresyan@mail.ru},
Gor Sarkissian$^{1,2}$
\footnote{sarkissn@theor.jinr.ru, gor.sargsyan@yerphi.am}
 and Vyacheslav P. Spiridonov $^{2,3}$\footnote{spiridon@theor.jinr.ru}
\end{center}
\begin{center}
$^1$Yerevan Physics Institute,\\
Alikhanian Br. 2, 0036\, Yerevan, Armenia
\end{center}
\begin{center}
$^2$ Bogoliubov Laboratory of Theoretical Physics, JINR,\\
Dubna, Moscow region, 141980 Russia\\
\end{center}
\begin{center}
$^3$ National Research University Higher School of Economics, Moscow, Russia\\
\end{center}

\vskip 1.5em

\begin{abstract} \noindent
We study properties of a parafermionic generalization of the hyperbolic hypergeometric function appearing as the most important part in the fusion matrix for Liouville field theory and the Racah-Wigner symbols for the
Faddeev modular double. We show that this  generalized hypergeometric function is a limiting form of the rarefied elliptic hypergeometric function $V^{(r)}$ and derive its transformation properties and
a mixed difference-recurrence equation satisfied by it. At the intermediate level we describe symmetries
of  a more general rarefied hyperbolic hypergeometric function.
An important $r=2$ case corresponds to the supersymmetric hypergeometric function given by the integral
appearing in the fusion matrix of $N=1$ super Liouville field theory and the Racah-Wigner symbols of the quantum algebra ${\rm U}_q({\rm osp}(1|2))$. We indicate relations to the standard Regge symmetry and
prove some previous conjectures for the supersymmetric Racah-Wigner symbols by establishing
their different parametrizations.
\end{abstract}


\tableofcontents


\section{Introduction}

It is well known that in order to solve some problems in physics one should first master the corresponding
mathematical tools. There are now many topics in modern mathematical physics where properties of the elliptic
gamma functions \cite{ruij,spi:theta} and  the Faddeev modular quantum dilogarithm \cite{Fad94,Fad95}
(or the hyperbolic gamma function) play an important role. In particular, they are needed in the investigations
of $4D$ $N=1$ superconformal field theories \cite{DO}, $3D$ $N=2$ supersymmetric field theories \cite{will},
Liouville field theory \cite{Teschner:2001rv}, Ruijsenaars-Schneider and van Diejen integrable models
\cite{ruijj,vD,spi:tmf}, etc.
In all these applications a crucial role is played by integrals of various products of these
generalized gamma functions, which are called elliptic/hyperbolic hypergeometric integrals.
In the case of $4D$ $N=1$ supersymmetric theories elliptic hypergeometric integrals \cite{spi:umn,spi:essays}
define superconformal indices \cite{DO}, and in the case of $3D$ $N=2$ supersymmetric theories hyperbolic hypergeometric integrals define partition functions  \cite{will}.

For the Liouville field theory and Ruijsenaars-Schneider model a special integral of
a product of eight hyperbolic gamma functions (given by formula (\ref{jhhh}) below)
introduced by  Ruijsenaars \cite{ruijj} is particularly important.
It enters as the most essential part in the fusion matrix of the former model and gives eigenfunctions
of the Hamiltonian in the latter case. From the group-theoretical point of view, it describes
$6j$-symbols for the Faddeev modular double derived by Ponsot and Teschner \cite{Ponsot:2000mt}.
Various generalizations of the mentioned problems, such as studies of $4D$ $N=1$ supersymmetric  field theories
on the $L(r,1)\times S^1$ space-time, where $L(r,1)$ is the simplest lens space \cite{razwil},
$3D$ $N=2$ field theories on  $L(r,1)$ \cite{will}, and supersymmetric \cite{Hadasz:2007wi}
or parafermionic extensions of the Liouville field theory\cite{Bershtein:2010wz}, require consideration of
the so-called rarefied or lens elliptic \cite{razwil,Spiridonov:2016uae} and hyperbolic \cite{dimofte,Gahramanov:2016ilb,IY,SS,SS24} gamma functions and corresponding generalized hypergeometric functions.

In this paper we study a parafermionic generalization of the  Ruijsenaars hyperbolic hypergeometric function
(see expression (\ref{jparker}) below) given by a sum of integrals of the product of eight
rarefied hyperbolic gamma functions, as defined in \cite{SS}.
We show that this function can be derived as a degeneration of the rarefied elliptic hypergeometric $V^{(r)}$-function introduced in \cite{Spiridonov:2016uae}.
Applying a specific limiting procedure to the transformations of the  $V^{(r)}$-function found in \cite{Spiridonov:2016uae}, we deduce various symmetry properties of the parafermionic  hypergeometric
function and show that they can be considered as generalizations of the well
known Regge symmetries of $6j$-symbols for the $SU(2)$ group.

We consider in detail the supersymmetric case corresponding to $r=2$.
Specializing the so-called second symmetry transformation to $r=2$, we derive in a straightforward way
a supersymmetric analogue of the formula obtained earlier for the Racah-Wigner symbols of the Faddeev modular double \cite{tvar}.  After that we analyze  the supersymmetric
Racah-Wigner symbols suggested in \cite{Hadasz:2013bwa,Pawelkiewicz:2013wga}. First we check that corresponding
expressions indeed represent particular cases of the supersymmetric  hypergeometric function.
In particular, we show that all the weird sign factors that appear in these expressions come simply
from the change of signs in the definition of the rarefied hyperbolic gamma functions entering
the supersymmetric  hypergeometric function (see equation (\ref{l2s22})).
Then we check that the restriction to the Neveu-Schwarz sector of
the general supersymmetric Racah-Wigner symbol, suggested in \cite{Pawelkiewicz:2013wga},  is
in agreement with the  Neveu-Schwarz sector Racah-Wigner symbols derived in
\cite{Hadasz:2013bwa}.
And, finally, we give general expression for the supersymmetric Racah-Wigner symbols, considered in \cite{Pawelkiewicz:2013wga}, in all available parametrizations for both the Neveu-Schwarz and Ramond sectors.

The paper is organized as follows. In Sect. 2, we describe necessary properties of the rarefied
hyperbolic gamma function. In Sect. 3, we deduce transformation properties of the  hyperbolic
hypergeometric integral entering $6j$-symbols of the Faddeev modular double from the symmetry
properties of the elliptic $V$-function. In Sect. 4, we review difference-recurrence equations
for hyperbolic hypergeometric integrals. In Sect. 5, we define the parafermionic hypergeometric
integral and derive its symmetries from the corresponding transformation properties of the
rarefied elliptic $V^{(r)}$-function, whereas in Sect. 6 we obtain the difference-recurrence
equation for it. In Sect. 7 we study in detail  the supersymmetric case.
Appendices contain some auxiliary material.

\section{Properties of the rarefied elliptic and hyperbolic gamma functions }

The standard elliptic gamma function $\Gamma(z;p,q)$ can be defined as a double infinite product:
\be
\Gamma(z;p,q)=\prod_{j,k=0}^{\infty}{1-z^{-1}p^{j+1}q^{k+1}\over 1-zp^jq^k}\, ,
\quad |p|,|q|<1\, ,\quad z\in \mathbb{C}^*.
\ee
It is symmetric in bases $p$ and $q$, $\Gamma(z;p,q)=\Gamma(z;q,p)$, and satisfies the equations
$$
\Gamma(qz;p,q)=\theta(z;p)\Gamma(z;q,p), \qquad
\Gamma(pz;p,q)=\theta(z;q)\Gamma(z;q,p),
$$
where $\theta(z;p)$ is a short Jacobi theta-function
$$
\theta(z;q)=(z;q)_\infty (qz^{-1};q)_\infty, \quad (z;q)_\infty:=\prod_{j=0}^\infty (1-zq^j),
$$
which is related to the standard Jacobi $\theta_1$-function as follows
\begin{eqnarray} \nonumber &&
 \theta_1(u|\tau)=-\theta_{11}(u)=-\sum_{\ell\in \mathbb Z+1/2}e^{\pi \textup{i} \tau \ell^2}
e^{2\pi \textup{i} \ell (u+1/2)}
\\  && \makebox[4em]{}
=\textup{i}q^{1/8}e^{-\pi \textup{i} u}(q;q)_\infty\theta(e^{2\pi \textup{i}u};q),
\qquad q=e^{2\pi \textup{i}\tau}\, .
\nonumber\end{eqnarray}

The elliptic gamma function associated with the simplest lens space is defined as a product
of two standard elliptic gamma functions with different bases  \cite{razwil},
\bea\label{lensf}
&&\gamma_e(z,m; p,q)=\Gamma(zp^m; p^r,pq)\Gamma(zq^{r-m};q^r,pq)\\ \nonumber
&& \makebox[2em]{}
=\prod^{\infty}_{j,k=0}{1-z^{-1}p^{-m}(pq)^{j+1}p^{r(k+1)}\over 1-zp^m(pq)^jp^{rk}}
{1-z^{-1}q^m(pq)^{j+1}q^{rk}\over 1-zq^{r-m}(pq)^jq^{rk}}\, ,
\quad m\in\mathbb{Z}.
\eea
As shown in  \cite{Spiridonov:2016uae}, function (\ref{lensf}) can be written as a special
product of the standard elliptic gamma functions
with bases $p^r$ and $q^r$. For the fundamental region $0\leq m\leq r$ it has the form:
$$
\gamma_e(z,m; p,q)=\prod_{k=0}^{m-1}\Gamma(q^{r-m}z(pq)^k;p^r,q^r)\prod_{k=0}^{r-m-1}\Gamma(p^mz(pq)^k;p^r,q^r)\, .
$$
The quasiperiodicity property:
\be\label{quasyper}
{\gamma_e(z,m+kr;p,q)\over \gamma_e(z,m;p,q)}=\left(-{\sqrt{pq}\over z}\right)^{mk+r{k(k-1)\over 2}}
\left({q\over p}\right)^{k\left({m^2\over 2}+mr{k-1\over 2}+r^2{(k-1)(2k-1)\over 12}\right)},\quad k\in \mathbb{Z},
\ee
established in \cite{Spiridonov:2016uae}, implies, that for all other values of $m$, the function  $\gamma_e(z,m; p,q)$ is determined by its form in the fundamental interval of $m$.
For further use, it is also convenient to introduce the function:
\bea\label{gr}
&&\Gamma^{(r)}(z,m; p,q) =
(-z)^{m(m-1)\over 2}p^{m(m-1)(m-2)\over 6}q^{-{m(m-1)(m+1)\over 6}}\gamma_e(z,m;p,q)\, .
\eea
The elliptic gamma function has the following asymptotic behaviour \cite{ruij,rai:limits}:
\be\label{ruijik}
\Gamma(e^{-2\pi vy};e^{-2\pi v\omega_1}, e^{-2\pi v\omega_2})
\stackreb{=}{ v\to 0}
e^{-\pi(2y-\omega_1-\omega_2)/12v\omega_1\omega_2}\gamma^{(2)}(y;\omega_1,\omega_2),
\ee
where $\gamma^{(2)}(y;\omega_1,\omega_2)$ is the Faddeev modular quantum dilogarithm
\cite{Fad94,Fad95} or the hyperbolic gamma function \cite{ruij, ruijj}.
The parameter $v$ approaches $0$ along the positive real axis
and parameters $\omega_1$ and $\omega_2$ have positive real parts,  ${\rm Re}(\omega_{1,2})>0$.

We use the shorthand notation $\gamma^{(2)}(y;\mathbf{\omega}):=\gamma^{(2)}(y;\omega_1,\omega_2)$
when the quasiperiods coincide with $\omega_1$ and $\omega_2$. Otherwise the quasiperiods will be written
explicitly. The function $\gamma^{(2)}(y;\mathbf{\omega})$ has the integral representation
\be
\gamma^{(2)}(y;\mathbf{\omega})=\exp\left(-\int_0^{\infty}\left({\sinh(2y-\omega_1-\omega_2)x\over 2\sinh(\omega_1x)
\sinh(\omega_2x)}-{2y-\omega_1-\omega_2\over 2\omega_1\omega_2x}\right)\right){dx\over x}
\ee
and obeys the equations:
\be\label{hp1}
{\gamma^{(2)}(y+\omega_1;\mathbf{\omega})\over \gamma^{(2)}(y;\mathbf{\omega})}=2\sin{\pi y\over \omega_2}\,  ,\quad
{\gamma^{(2)}(y+\omega_2;\mathbf{\omega})\over \gamma^{(2)}(y;\mathbf{\omega})}=2\sin{\pi y\over \omega_1}.
\ee
It has the following
asymptotics \cite{Kharchev:2001rs}:
\bea\nonumber
&&\stackreb{\lim}{y\to \infty}\gamma^{(2)}(y;\mathbf{\omega})=e^{-{\textup{i}\pi\over 2}B_{2,2}(y;\mathbf{\omega})},
\quad {\rm for}\; {\rm arg}\;\omega_1<{\rm arg}\; y<{\rm arg}\;\omega_2+\pi,
\\ &&
\stackreb{\lim}{y\to \infty}\gamma^{(2)}(y;\mathbf{\omega})=e^{{\textup{i}\pi\over 2}B_{2,2}(y;\mathbf{\omega})},
\quad {\rm for}\; {\rm arg}\;\omega_1-\pi<{\rm arg}\; y<{\rm arg}\;\omega_2,
\label{asy1}\eea
where $B_{2,2}(y;\mathbf{\omega})$  is the second order Bernoulli polynomial:
$$
B_{2,2}(y;\mathbf{\omega})=\frac{1}{\omega_1\omega_2}
\left((y-\frac{\omega_1+\omega_2}{2})^2-\frac{\omega_1^2+\omega_2^2}{12}\right).
$$

Using asymptotics (\ref{ruijik}) one can show that \cite{SS}:
\be\label{limg}
\gamma_e\left(e^{-{2\pi vy\over r}}, m;e^{-{2\pi v\omega_1\over r}},
e^{-{2\pi v\omega_2\over r}}\right)\stackreb{=}{ v\to 0}e^{-\pi(2y-\omega_1-\omega_2)/12v\omega_1\omega_2}\Lambda(y;m;\mathbf{\omega})\, ,
\ee
where the function $\Lambda(y,m;\mathbf{\omega})$ is defined as follows.
For the fundamental region $0\leq m\leq r$ one has
\bea \nonumber &&
\Lambda(y,m,\mathbf{\omega})=\prod_{k=0}^{m-1}\gamma^{(2)}
\left({y\over r}+\omega_2\left(1-{m\over r}\right)+(\omega_1+\omega_2){k\over r};\mathbf{\omega}\right)
\\ && \makebox[2em]{} \times
\prod_{k=0}^{r-m-1}\gamma^{(2)}
\left({y\over r}+{m\over r}\omega_1+(\omega_1+\omega_2){k\over r};\mathbf{\omega}\right)\, ,
\label{Lambda1}\eea
Qusiperiodicity relation (\ref{quasyper}) in  the limit (\ref{limg}) yields
\be\label{lambdasign}
\Lambda(y,m+kr;\mathbf{\omega})=(-1)^{mk+r{k(k-1)\over 2}}\Lambda(y,m;\mathbf{\omega}).
\ee
As we see, the function $\Lambda(y,m;\mathbf{\omega})$ for all the  values of $m$ is determined by its form in the fundamental region.
We also have from the asymptotics (\ref{limg}) and definition (\ref{gr}):
\be\label{limgg}
\Gamma^{(r)}\left(e^{-{2\pi vy\over r}}, m;e^{-{2\pi v\omega_1\over r}},
e^{-{2\pi v\omega_2\over r}}\right)\stackreb{=}{ v\to 0}e^{-\pi(2y-\omega_1-\omega_2)/12v\omega_1\omega_2}(-1)^{m(m-1)\over 2}\Lambda(y;m;\mathbf{\omega}).
\ee
Applying the limit (\ref{limg}) to definition (\ref{lensf}) one can
derive another expression for the function $\Lambda(y,m;\mathbf{\omega})$:
\be
\Lambda(y;m;\mathbf{\omega})=\gamma^{(2)}\left({y+m\omega_1\over r};\omega_1,{\omega_1+\omega_2\over r}\right)
\gamma^{(2)}\left({y+(r-m)\omega_2\over r};\omega_2,{\omega_1+\omega_2\over r}\right).
\label{Lambda2}\ee
Using this expression one can show that the function $\Lambda(y;m;\mathbf{\omega})$ has the following
asymptotics \cite{Nieri:2015yia,SS}:
\bea \nonumber &&
\stackreb{\lim}{y\to \infty}\Lambda(y;m;\mathbf{\omega})=e^{-{\textup{i}\pi\over 2}\left({1\over r}B_{2,2}(y;\mathbf{\omega})+{m^2\over r}-m+{r\over 6}-{1\over 6r}\right)},
\quad {\rm for}\; {\rm arg}\;\omega_1<{\rm arg}\; y<{\rm arg}\;\omega_2+\pi,
\\ &&
\stackreb{\lim}{y\to \infty}\Lambda(y;m;\mathbf{\omega})=e^{{\textup{i}\pi\over 2}\left({1\over r}B_{2,2}(y;\mathbf{\omega})+{m^2\over r}-m+{r\over 6}-{1\over 6r}\right)},
\quad {\rm for}\; {\rm arg}\;\omega_1-\pi<{\rm arg}\; y<{\rm arg}\;\omega_2.
 \label{asy1'}\eea

We also would like to note that below we use the following shorthand notation:
\bea\label{shorts}
&&\Gamma(az^{\pm k}; p,q)\equiv\Gamma(az^k; p,q)\Gamma(az^{-k}; p,q),\nonumber\\
&&\Gamma^{(r)}(az^{\pm k}, m\pm n ;p,q)\equiv\Gamma^{(r)}(az^k, m+n; p,q)\Gamma^{(r)}(az^{-k},m-n; p,q),\nonumber\\
&&\gamma^{(2)}(y\pm x;\mathbf{\omega})\equiv\gamma^{(2)}(y+x;\mathbf{\omega})\gamma^{(2)}(y-x;\mathbf{\omega}),\nonumber\\
&&\Lambda(y\pm x ;m\pm n;\mathbf{\omega})\equiv\Lambda(y+ x ;m+ n;\mathbf{\omega})\Lambda(y- x ;m- n;\mathbf{\omega}),\nonumber\\
&&\gamma^{(2)}(y,x;\mathbf{\omega})\equiv\gamma^{(2)}(y;\mathbf{\omega})\gamma^{(2)}(x;\mathbf{\omega}),\nonumber\\
&&\theta(x,y,z;p)\equiv\theta(x;p)\theta(y;p)\theta(z;p).
\eea

\section{Symmetries of $6j$-symbols for the Faddeev modular double}

In this section we review known symmetry properties of the following hyperbolic hypergeometric function
introduced by Ruijsenaars in 1994 \cite{ruij}
\be\label{jhhh}
J_h(\underline{\beta},\underline{\gamma})=\int_{-\textup{i}\infty}^{\textup{i}\infty}
\prod_{a=1}^4\gamma^{(2)}(\beta_a- z;\mathbf{\omega})
\gamma^{(2)}(\gamma_a+ z;\mathbf{\omega})\frac{dz}{\textup{i}\sqrt{\omega_1\omega_2}}
\ee
with the parameters $\beta_a$, $\gamma_a$ satisfying the balancing condition
\be\label{balk}
\sum_{a=1}^4(\gamma_a+\beta_a)=2(\omega_1+\omega_2),
\ee
and explain how they emerge from symmetries of the elliptic analogue of the Euler--Gauss hypergeometric function.
The contour of integration in \eqref{jhhh} is of the Mellin-Barnes type. Namely, it separates the sequences of poles
going to infinity in the right half-plane from those lying in the left half-plane.
For generic values of $\beta_a$ and  $\gamma_a$ such a contour always exists. With the conventional choice
$\omega_1=b$ and $\omega_2=b^{-1}$, the function \eqref{jhhh} defines a key element in the construction
of $6j$-symbols of the Faddeev modular quantum double and fusion matrix of the Liouville field theory
with the central charge $c=1+6(b+b^{-1})^2$ \cite{Ponsot:2000mt,Teschner:2001rv}.

\subsection{Symmetry ${\rm I}_a$}

Consider the $V$-function, an elliptic analogue of the Euler--Gauss hypergeometric function introduced in \cite{spi:essays},
\begin{equation}
V(g_1,\ldots,g_8;p,q)=\frac{(p;p)_\infty(q;q)_\infty}{4\pi {\rm i}}\int_{\mathbb{T}}
{\prod\limits_{a=1}^8\Gamma\big(g_az^{\pm1}; p,q\big)\over \Gamma\big(z^{\pm 2}; p,q\big)}{{\rm d}z\over z},
\end{equation}
where $\mathbb{T}$ is the unit circle with positive orientation, the parameters satisfy constraints $|g_a|<1$
and the balancing condition $ \prod\limits_{a=1}^8g_a=p^2q^2$. Here we used the shorthand notation \eqref{shorts}.
This function has the $W(E_7)$ Weyl group of  symmetry transformations, whose key generating
relation has been established in \cite{spi:theta}:
\begin{gather}\label{vt1}
V(g_1,\ldots,g_8;p,q)=\prod_{1\leq j< k\leq 4}\Gamma(g_jg_k;p,q)
\prod_{5\leq j< k\leq 8}\Gamma(g_jg_k;p,q) V(\tilde{g}_1,\ldots,\tilde{g}_8;p,q),
\end{gather}
where
\begin{equation}\nonumber
\tilde{g}_j=\rho^{-1}g_j,\qquad \tilde{g}_{j+4}=\rho g_{j+4},\quad j=1,2,3,4, \qquad
\rho=\sqrt{g_1g_2g_3g_4\over pq}=\sqrt{pq\over g_5g_6g_7g_8},
\end{equation}
with $|\tilde g_a|<1$.

Introduce a hyperbolic hypergeometric function $I_h(\underline{s})$ defined by the integral
\begin{equation}\label{defih}
I_h(\underline{s})=\int_{-\textup{i}\infty}^{\textup{i}\infty}{\prod_{j=1}^8\gamma^{(2)}(s_j\pm z;\mathbf{\omega})\over
\gamma^{(2)}(\pm 2z;\mathbf{\omega})}\frac{dz}{2\textup{i}\sqrt{\omega_1\omega_2}},
\end{equation}
with $s_j$ satisfying the conditions $\text{Re}(s_j)>0$ and
\begin{equation}\label{mu8}
\sum_{j=1}^8 s_j=2Q,\quad Q:=\omega_1+\omega_2
\end{equation}
(for the notation, see \eqref{shorts}). Applying the hyperbolic degeneration limit \eqref{ruijik} to the transformation rule \eqref{vt1}, one comes to the following identity \cite{BRS}

\begin{equation}\label{ide1}
I_h(\underline{s})=\prod_{1\leq j< k \leq 4}\gamma^{(2)}(s_j+s_k;\mathbf{\omega})
\prod_{5\leq j< k \leq 8}\gamma^{(2)}(s_j+s_k;\mathbf{\omega})\, I_h(\underline{\tilde{s}})\, ,
\end{equation}
where
$$
\tilde{s}_j=s_j+\eta, \quad \tilde{s}_{j+4}=s_{j+4}-\eta, \quad j=1,2,3,4,\quad
\eta={1\over 2}(\omega_1+\omega_2-\sum_{j=1}^4 s_j)
$$
with $\text{Re}(\tilde s_j)>0$. Let us parametrize
\be\label{limpro}
s_{1,2,5,6}=\gamma_{1,2,3,4}+\textup{i}\mu, \qquad
s_{3,4,7,8}=\beta_{1,2,3,4} - \textup{i}\mu
\ee
in (\ref{ide1}), shift the integration variable  $z\to z-\textup{i}\mu$ on both sides of the equality,
and take the limit $\mu\to -\infty$. Then we obtain \cite{BRS}:
\bea\label{ide1b}
&&J_h(\underline{\beta},\underline{\gamma})=\prod_{ j, k =1}^2\gamma^{(2)}(\beta_j+\gamma_k;\mathbf{\omega})\prod_{ j, k =3}^4\gamma^{(2)}(\beta_j+\gamma_k;\mathbf{\omega})
\\ \nonumber && \makebox[2em]{}
\times J_h(\beta_1+\eta,\beta_2+\eta,\beta_3-\eta,\beta_4-\eta,\gamma_1+\eta,\gamma_2+\eta,\gamma_3-\eta,\gamma_4-\eta),
\eea
where $J_h(\underline{\beta},\underline{\gamma})$ is the Ruijsenaars function (\ref{jhhh}), (\ref{balk}),
and $\eta={1\over 2}(\omega_1+\omega_2-\beta_1-\beta_2-\gamma_1-\gamma_2).$

\subsection{Symmetry ${\rm I}_b$}
Define another hyperbolic hypergeometric integral:
\be\label{ehrho}
E_h(\underline{\rho})=\int_{-\textup{i}\infty}^{\textup{i}\infty}{\prod_{i=1}^6\gamma^{(2)}(\rho_i \pm z;\mathbf{\omega})\over
\gamma^{(2)}(\pm 2z;\mathbf{\omega})}\frac{dz}{2\textup{i}\sqrt{\omega_1\omega_2}}
\ee
(for notation, see \eqref{shorts}). Applying in equality (\ref{ide1}) a different parametrization
$$
s_{1,2,3,8}=\beta_{1,2,3,4}+\textup{i}\mu, \qquad
s_{4,5,6,7}=\gamma_{4,1,2,3}-\textup{i}\mu,
$$
shifting  the integration variable $z\to z+\textup{i}\mu$ only in the left-hand side integral,
and taking the limit $\mu\to\infty$, one finds \cite{BRS}:
\bea \nonumber
&& J_h(\underline{\beta},\underline{\gamma})=\prod_{i=1}^3\gamma^{(2)}(\beta_i+\gamma_4;\mathbf{\omega})
\gamma^{(2)}(\gamma_i+\beta_4;\mathbf{\omega})
\\ && \makebox[2em]{} \times
E_h(\beta_1+\xi,\beta_2+\xi,\beta_3+\xi,\gamma_1-\xi,\gamma_2-\xi,\gamma_3-\xi),
\label{jheh}\eea
where $2\xi=Q-\gamma_4-\sum_{i=1}^3\beta_i.$

\subsection{Symmetry II}

The second type of identities follows from the key generating relation \eqref{vt1} after a group action composition
(a repetition of  \eqref{vt1} after a permutation of parameters),
\begin{equation}
V(g_1,\ldots,g_8;p,q)=\prod_{j, k=1}^4\Gamma(g_jg_{k+4};p,q)
V\Big(\tfrac{T^{1/2}}{g_1},\ldots,\tfrac{T^{1/2}}{g_4},\tfrac{U^{1/2}}{g_5},\ldots,\tfrac{U^{1/2}}{g_8};p,q\Big),
\label{vt2}\end{equation}
where $T=g_1g_2g_3g_4$, $U=g_5g_6g_7g_8$, and $|T^{1/2}/g_j|, |U^{1/2}/g_{j+4}|<1,\, j=1,\ldots, 4$.
The hyperbolic degeneration limit~\eqref{ruijik} for integrals described in the previous sections
reduces relation \eqref{vt2} to the following identity for $I_h(\underline{s})$ function \cite{Sarkissian:2020ipg}:
\begin{gather}\label{ide2}
I_h(\underline{s})=\prod_{j,k=1}^4\gamma^{(2)}(s_j+s_{k+4};\mathbb{\omega})
I_h(G-s_1,\ldots,G-s_4,Q-G-s_5,\ldots,Q-G-s_8),
\end{gather}
where $G:={1\over 2}\sum\limits_{j=1}^4 s_j$ and $Q=\omega_1+\omega_2$.
Now, for the parametrization (\ref{limpro}), the $\mu\to \infty$ limit  yields
\bea \nonumber
&&
J_h(\underline{\beta},\underline{\gamma})=\prod_{j,k=1}^2\gamma^{(2)}(\gamma_j+\beta_{k+2};\mathbf{\omega})
\gamma^{(2)}(\gamma_{j+2}+\beta_k;\mathbf{\omega})J_h(G-\gamma_1,G-\gamma_2,
\\   \label{newsymka} && \makebox[1em]{}
Q-G-\gamma_3,Q-G-\gamma_4; G-\beta_1,G-\beta_2,Q-G-\beta_3,Q-G-\beta_4),
\eea
where $G={1\over 2}(\gamma_1+\gamma_2+\beta_1+\beta_2)$.

Recall now the Ponsot-Teschner formula \cite{Ponsot:2000mt} for $6j$-symbols of the Faddeev modular quantum double
$U_q(sl(2,\mathbb{R})\otimes U_{\tilde{q}}(sl(2,\mathbb{R})$, $q=e^{\pi \textup{i}b^2}$ and $\tilde{q}=e^{\pi \textup{i}b^{-2}}$:
\be\label{6jpt}
\big\{\,{}^{\alpha_1}_{\alpha_3}\,{}^{\alpha_2}_{{\alpha}_4}\,\big|\,{}^{\alpha_s}_{\alpha_t}\big\}_b=
{S_b(\alpha_s+\alpha_2-\alpha_1)S_b(\alpha_1+\alpha_t-\alpha_4)\over
S_b(\alpha_t+\alpha_2-\alpha_3)S_b(\alpha_3+\alpha_s-\alpha_4)}
|S_b(2\alpha_t)|^2J_h(\beta_a^{\circ},\gamma_a^{\circ};b),
\ee
where
$$
J_h(\beta_a^{\circ},\gamma_a^{\circ}; b)=\int_{-\textup{i}\infty}^{\textup{i}\infty}\prod_{a=1}^4S_b(z+\gamma_a^{\circ})
S_b(-z+\beta_a^{\circ})
\frac{dz}{\textup{i}},
\qquad
S_b(x):=\gamma^{(2)}(x,b,b^{-1}),
$$
and  $\gamma_a^{\circ},\beta_a^{\circ},\; a=1,\ldots,4$, are the Ponsot-Teschner parameters:
\be\label{ptparam}
\begin{aligned}
&\gamma_1^{\circ}=-Q/2+\alpha_3-\alpha_4\, ,\\
&\gamma_2^{\circ}=-Q/2+\alpha_1-\alpha_2\, ,\\
&\gamma_3^{\circ}=Q/2-\alpha_3-\alpha_4\, ,\\
&\gamma_4^{\circ}=Q/2-\alpha_1-\alpha_2\, ,
\end{aligned}\qquad
\begin{aligned}
&\beta_1^{\circ}=Q/2+\alpha_s\, ,\\
&\beta_2^{\circ}=Q/2-\alpha_t+\alpha_4+\alpha_2\, ,\\
&\beta_3^{\circ}=-Q/2+\alpha_t+\alpha_4+\alpha_2\, ,\\
&\beta_4^{\circ}=3Q/2-\alpha_s\, .
\end{aligned}
\ee

Expression \eqref{6jpt} gives also, up to some normalization factors, the fusion matrix of the Liouville field theory with the central charge $c=1+6Q^2$. In fact, in this form the parameters \eqref{ptparam} appeared in \cite{tvar}, but they can be easily derived from the ones used in
\cite{Ponsot:2000mt} after shifting the integration variable by $-Q/2-\alpha_s$.
Here and below when we shift the integration variable in such a way, we assume that the integration contour is
deformed in an appropriate way.

Using relation  $2G=\alpha_s-\alpha_t+\alpha_1+\alpha_3$
and also shifting the integration variable $z$ by $G+Q+\beta_3^{\circ}$ in the integral standing on the right-hand side of relation \eqref{newsymka},  we obtain the identity
\bea\label{tvfor}
J_h(\underline{\beta}^{\circ},\underline{\gamma}^{\circ})=\Omega(\underline{\alpha})J_h(\underline{\beta}^{\diamond},\underline{\gamma}^{\diamond})\,,
\eea
where
\be\label{tvparam}
\begin{aligned}
&\gamma_1^{\diamond}=\alpha_{1234}\, ,\\
&\gamma_2^{\diamond}=\alpha_{13st}\, ,
\end{aligned}\qquad
\begin{aligned}
&\gamma_3^{\diamond}=2Q\, ,\\
&\gamma_4^{\diamond}=\alpha_{24st}\, ,
\end{aligned}\qquad
\begin{aligned}
&\beta_1^{\diamond}=-\alpha_{23t}\, ,\\
&\beta_2^{\diamond}=-\alpha_{14t}\, ,
\end{aligned}\qquad
\begin{aligned}
&\beta_3^{\diamond}=-\alpha_{12s}\, ,\\
&\beta_4^{\diamond}=-\alpha_{34s}\, ,
\end{aligned}
\ee
$$
\alpha_{ijk}\equiv \alpha_i+\alpha_j+\alpha_k\, ,\quad \alpha_{ijkl}\equiv \alpha_i+\alpha_j+\alpha_k+\alpha_l
$$
and
\bea \nonumber && \Omega(\underline{\alpha})=\gamma^{(2)}(Q+\alpha_s-\alpha_3-\alpha_4;\mathbf{\omega})
\gamma^{(2)}(Q+\alpha_s-\alpha_1-\alpha_2;\mathbf{\omega})
\\ \nonumber  && \makebox[2em]{} \times \gamma^{(2)}(Q-\alpha_t+\alpha_2-\alpha_3;\mathbf{\omega})\gamma^{(2)}(Q-\alpha_t+\alpha_4-\alpha_1;\mathbf{\omega})
\\ \nonumber && \makebox[2em]{} \times \gamma^{(2)}(-Q+\alpha_t+\alpha_2+\alpha_3;\mathbf{\omega})\gamma^{(2)}(Q-\alpha_s+\alpha_3-\alpha_4;\mathbf{\omega})
\\ && \makebox[2em]{}  \times
\gamma^{(2)}(-Q+\alpha_t+\alpha_4+\alpha_1;\mathbf{\omega})
\gamma^{(2)}(Q-\alpha_s+\alpha_1-\alpha_2;\mathbf{\omega}).
\label{shortnot}\eea
Equality (\ref{tvfor}) was derived in  \cite{tvar} in a substantially more complicated way, and it was used
for finding of the hyperbolic volume of a non-ideal tetrahedron in the quasi-classical limit of $6j$-symbols
for the Faddeev modular double.

\subsection{Symmetry III}

The third form of the symmetry transformation for the $V$-function follows from equating right-hand
side expressions in \eqref{vt1} and \eqref{vt2},
\begin{gather}
V(g_1,\ldots,g_8;p,q)=\prod_{1\leq j< k\leq 8}\Gamma(g_jg_{k};p,q)
V\left(\frac{\sqrt{pq}}{g_1},\ldots,\frac{\sqrt{pq}}{g_8};p,q\right).
\end{gather}
The hyperbolic degeneration limit (\ref{ruijik}) results in the following symmetry transformation
for the $I_h(\underline{s})$ function \cite{BRS}

\begin{equation}\label{reflt}
I_h(\underline{s})=\prod_{1\leq j< k\leq 8}\gamma^{(2)}(s_j+s_k;\mathbf{\omega})
I_h\left(\underline{\lambda}\right), \qquad \lambda_j =\frac{\omega_1+\omega_2}{2}-s_j.
\end{equation}

For reducing this relation, we replace in (\ref{reflt})
\[
s_j \to \gamma_j+{\rm i}\mu, \qquad s_{j+4}=\beta_j-{\rm i}\mu, \qquad j=1,\ldots, 4,
\qquad z\to z-{\rm i}\mu,
\]
where $z$ denotes integration variable on both sides of the equality.
The balancing condition takes the form $\sum\limits_{j=1}^4 (\beta_j+\gamma_j)=2Q$.
After going to the limit $\mu\to -\infty$, we come to the identity \cite{Sarkissian:2020ipg}
\begin{gather}\nonumber
\int_{-{\rm i}\infty}^{{\rm i}\infty}\prod_{j=1}^4\gamma^{(2)}(\gamma_j+z;\mathbf{\omega})
\gamma^{(2)}(\beta_j-z;\mathbf{\omega}) {\rm d}z
=\prod_{j,k=1}^4 \gamma^{(2)}(\gamma_j+\beta_k;\mathbf{\omega})
\\ \qquad{} \times
\int_{-{\rm i}\infty}^{{\rm i}\infty}\prod_{j=1}^4\gamma^{(2)}\big(\tfrac{1}{2}Q-\beta_j+z;\mathbf{\omega}\big)
\gamma^{(2)}(\tfrac{1}{2}Q-\gamma_j-z;\mathbf{\omega}) {\rm d}z.
\label{infy} \end{gather}

\subsection{Relation to the Regge symmetry}

The standard Regge symmetry for the $SU(2)$ group Racah-Wigner symbols corresponds to
a particular reflection transformation invariance
\cite{Regge:1959ze}:
\be\label{regsym}
\big\{\,{}^{\alpha_1}_{\alpha_3}\,{}^{\alpha_2}_{\alpha_4}\,\big|\,{}^{\alpha_s}_{\alpha_t}\big\}=
\big\{\,{}^{S-\alpha_1}_{S-\alpha_3}\,{}^{S-\alpha_2}_{S-\alpha_4}\,\big|\,{}^{\alpha_s}_{\alpha_t}\big\}\,,
\ee
where $S={1\over 2}(\alpha_1+\alpha_2+\alpha_3+\alpha_4).$ Consider the effect of this reflection
of $\alpha_i, \, i=1,\ldots,4,$ for the parameters entering Ponsot-Teschner $6j$-symbols
in both the original definition \eqref{6jpt} and the transformation \eqref{tvfor}.

The transformation $\alpha_i \to S -\alpha_i$ brings parameters \eqref{ptparam} to the form
\be\label{ptparamr}
\begin{aligned}
&\gamma_1^{r}=-Q/2-\alpha_3+\alpha_4\, ,\\
&\gamma_2^{r}=-Q/2-\alpha_1+\alpha_2\, ,\\
&\gamma_3^{r}=Q/2-\alpha_1-\alpha_2\, ,\\
&\gamma_4^{r}=Q/2-\alpha_3-\alpha_4\, ,
\end{aligned}\qquad
\begin{aligned}
&\beta_1^{r}=Q/2+\alpha_s\, ,\\
&\beta_2^{r}=Q/2-\alpha_t+\alpha_1+\alpha_3\, ,\\
&\beta_3^{r}=-Q/2+\alpha_t+\alpha_1+\alpha_3\, ,\\
&\beta_4^{r}=3Q/2-\alpha_s\, .
\end{aligned}
\ee
The same form of parameters follows from the permutation symmetry and transformation ${\rm I}_a$ \eqref{ide1b}.
Indeed, shifting the integration variable for $J_h$ in the right-hand side of \eqref{ide1b} by $\eta$,
we can write
\bea \nonumber
&& J_h(\beta_1+\eta,\beta_2+\eta,\beta_3-\eta,\beta_4-\eta,\gamma_1+\eta,\gamma_2+\eta,\gamma_3-\eta,\gamma_4-\eta)
\\ && \makebox[2em]{}
= J_h(\beta_1+2\eta,\beta_2+2\eta,\beta_3,\beta_4,\gamma_1,\gamma_2,\gamma_3-2\eta,\gamma_4-2\eta).
\label{ide1bb}\eea
Since the left-hand side expression in  \eqref{ide1b} is symmetric in $\beta_a$ and $\gamma_a$, we can permute
indices of the latter parameters in the right-hand side as well. If we take a partner of
\eqref{ide1b} with the following function on the right-hand side
$$
J_h(\beta_1^{\circ},\beta_2^{\circ}+2\eta,\beta_3^{\circ}+2\eta,\beta_4^{\circ},\gamma_1^{\circ}-2\eta,\gamma_2^{\circ}-2\eta,\gamma_3^{\circ},\gamma_4^{\circ})
$$
where $2\eta=Q-(\beta_2^{\circ}+\beta_3^{\circ}+\gamma_3^{\circ}+\gamma_4^{\circ})
=\alpha_1-\alpha_2+\alpha_3-\alpha_4$,
we obtain parameters \eqref{ptparamr} in a permuted order.

If we apply Regge reflection to parameters \eqref{tvparam},
which we recall were obtained from \eqref{ptparam} by the second symmetry transformation,
we just permute them with each other. Therefore the latter parametrization has the advantage to make  $J_h(\underline{\beta}^{\diamond},\underline{\gamma}^{\diamond})$ explicitly Regge invariant.
One can check, that in this parametrization the prefactors of $J_h$-function in the expression
for $6j$-symbols are also invariant under this transformation. In the original parametrization
\eqref{ptparam} the invariance of $6j$-symbols follows from the nontrivial additional
$S_b$-function factors emerging in the transformation  \eqref{ide1b}.

This analysis shows that we can consider symmetries of the generalized hypergeometric functions
as extensions of the Regge and permutation group symmetries of $6j$-symbols for the $SU(2)$ group
to the $6j$-symbols of the Faddeev modular quantum double. It is interesting to note that from the
latter $6j$-symbols one can obtain by various limiting procedures
$6j$-symbols of the unitary principal series representations of the $SL(2,\mathbb{C})$ group \cite{Derkachov:2021thp}, and of the $U_q(su(2))$ quantum group \cite{Pawelkiewicz:2013wga}.
So, the work \cite{Derkachov:2021thp} contains a generalization of the Regge and reflection symmetries
to the $6j$-symbols for the unitary principal series representations of the $SL(2,\mathbb{C})$ group.
Taking particular limits one can rederive symmetries for the $6j$-symbols of  $U_q(su(2))$ quantum
group and the classical $SU(2)$ group. In the last section, we obtain analogues of the Regge
and reflection symmetries for $6j$-symbols of the quantum supergroup ${\rm U}_q({\rm osp}(1|2))$.

\section{Difference equations}

As shown in \cite{spi:thesis,spi:tmf}, the $V$-function satisfies the following finite-difference equation
called the elliptic hypergeometric equation
\be\label{elldif}
{\mathcal L}(\underline{g})(U(qg_6,q^{-1}g_7)-U(g))+(g_6\leftrightarrow g_7)+U(g)=0,
\ee
where
$$
U(\underline{g})={V(g_1,\ldots,g_8;p,q)\over \Gamma(g_6 g_8^{\pm 1}; p,q)\Gamma(g_7 g_8^{\pm 1}; p,q)}
$$
and
\bea\nonumber
{\mathcal L}(\underline{g})={\theta\left({g_6\over qg_8};p\right)\theta\left(g_6 g_8; p\right)\theta\left({g_8\over g_6};p\right)
\over \theta\left({g_6\over g_7}; p\right)\theta\left({g_7\over qg_6};p\right)\theta\left({g_7g_6\over q};p\right)}
\prod_{k=1}^5{\theta\left({g_7g_k\over q};p\right)\over \theta(g_8g_k;p)}.
\eea
In \eqref{elldif} $(g_6\leftrightarrow g_7)$ means that there stands the previous expression with the parameters $g_6$
and $g_7$ permuted.

Using the asymptotic relations (\ref{ruijik}) and
\be\label{limthet}
\theta(e^{-2\pi  vy}; e^{-2\pi  v\omega_1})
\stackreb{=}{v\to 0} e^{-{\pi\over 6v\omega_1}}2\sin{\pi y\over \omega_1}\,,
\ee
one can derive the following difference equation for the corresponding hyperbolic hypergeometric
function \cite{BRS}:
\be\label{br}
{\mathcal A}(s;\omega_1,\omega_2)(Y(s_6+\omega_2,s_7-\omega_2)-Y(s))+(s_6\leftrightarrow s_7)+Y(s)=0,
\ee
where
\bea\label{ba'} && \nonumber
{\mathcal A}(s;\omega_1,\omega_2)={\sin{{\pi\over\omega_1}}(s_6-s_8-\omega_2)\sin{{\pi\over \omega_1}}(s_6+s_8)\sin{{\pi\over \omega_1}}(s_8-s_6)
\over \sin{{\pi\over \omega_1}}(s_6-s_7)\sin{{\pi\over \omega_1}}(s_7-s_6-\omega_2)\sin{{\pi\over \omega_1}}(s_7+s_6-\omega_2)}
\\ && \makebox[6em]{} \times
\prod_{k=1}^5{\sin{{\pi\over \omega_1}}(s_7+s_k-\omega_2)\over \sin{{\pi\over \omega_1}}(s_8+s_k)}
\eea
and
$$
Y(s)={I_h(\underline{s})\over \gamma^{(2)}(s_6\pm s_8, s_7\pm s_8; \mathbf{\omega})}.
$$

We now reparametrize $s_a$ in the identity (\ref{br}) in the following asymmetric way
$$
 s_a= \gamma_a+\textup{i}\mu, \quad s_{a+4}= \beta_{a}-\textup{i}\mu,  \quad a=1,2,3,4.
$$
Then the balancing condition takes the form
$ 
\sum_{a=1}^4(\gamma_a+\beta_a)=2(\omega_1+\omega_2).
$
Now we shift in (\ref{defih}) the integration variable $z\to z-\textup{i}\mu$ and take the limit
$\mu\to -\infty$. As a result, we obtain from the above equation
\be\label{secdif}
{\mathcal D}(\underline{\beta},\underline{\gamma};\omega_1,\omega_2)
(U(\beta_2+\omega_2,\beta_3-\omega_2)-U(\underline{\beta},\underline{\gamma}))+(\beta_2\leftrightarrow \beta_3)+U(\underline{\beta},\underline{\gamma})=0,
\ee
where
\bea \nonumber  &&
{\mathcal D}(\underline{\beta},\underline{\gamma};\omega_1,\omega_2)
={\sin{{\pi\over\omega_1}}(\beta_2-\beta_4-\omega_2)\sin{{\pi\over \omega_1}}(\beta_4-\beta_2)
\over \sin{{\pi\over \omega_1}}(\beta_2-\beta_3)\sin{{\pi\over \omega_1}}(\beta_3-\beta_2-\omega_2)}
\prod_{k=1}^4{\sin{{\pi\over \omega_1}}(\beta_3+\gamma_k-\omega_2)\over \sin{{\pi\over \omega_1}}(\beta_4+\gamma_k)},
\\ \nonumber && \makebox[8em]{}
 U(\underline{\beta},\underline{\gamma})={J_h(\underline{\beta},\underline{\gamma})
 \over \gamma^{(2)}(\beta_2- \beta_4, \beta_3- \beta_4; \mathbf{\omega})}.
\eea
Equation (\ref{secdif}) was derived in \cite{Derkachov:2021thp}.

\section{Symmetries of the parafermionic  hypergeometric integral}

For a positive integer $r$ we define the function
\bea\label{jparker} && \makebox[-2em]{}
J_{\epsilon}(\underline{\beta},\underline{l};\underline{\gamma},\underline{t})=\int_{-\textup{i}\infty}^{\textup{i}\infty}
\sum_{m\in \mathbb{Z}_r+\epsilon}
\prod_{a=1}^4\Lambda(-y+\beta_a;l_a-m;\mathbf{\omega})\prod_{a=1}^4\Lambda(y+\gamma_a;t_a+m;\mathbf{\omega})
{dy\over \textup{i}r\sqrt{\omega_1\omega_2}},
\eea
where the rarefied hyperbolic gamma function $\Lambda$ is defined in \eqref{Lambda1} or \eqref{Lambda2}.
Here $t_a\, , l_a\in \mathbb{Z}+\epsilon$ and $\epsilon=0,{1\over 2}$. Parameters $\gamma_j$, $\beta_j$ and $l_j$, $t_j$ satisfy the constraints:
\begin{equation}\label{mmm}
\sum_{j=1}^4 (\gamma_j+\beta_j)=2Q,\qquad \sum_{j=1}^4 (l_j+t_j)=0.
\end{equation}
Function (\ref{jparker}) is a parafermionic generalization of the integral (\ref{jhhh}).
In fact, the case when $\epsilon=1/2$ does not define new integral, since by the simultaneous
shifts $l_a\to l_a-\epsilon$,
$t_a\to t_a+\epsilon$, $m\to m-\epsilon$ one can eliminate the parameter $\epsilon$. But we keep it, because
it allows us in coming sections to write parafermionic generalizations of the symmetry relations described above
in a symmetric way. A relation of function \eqref{jparker} to the two-dimensional conformal field theory
is discussed below in section 7.

\subsection{Parafermionic symmetry ${\rm I}_a$}

Let us consider the rarefied elliptic hypergeometric function \cite{Spiridonov:2016uae}:
\be
V_\epsilon^{(r)}(\underline{g},\underline{n} ;p,q)=\frac{(p^r;p^r)_\infty (q^r;q^r)_\infty}{4\pi \textup{i}}
\sum_{m\in \mathbb{Z}_r+\epsilon}\int_{\mathbb T}
\frac {\prod_{a=1}^8\Gamma^{(r)} (g_az^{\pm 1},n_a\pm m;p,q)}
{\Gamma^{(r)} (z^{\pm 2},\pm 2m;p,q)} \frac{dz}{z},
\label{rfV}\ee
where $g_a\in \mathbb{C}^*, |g_a|<1$, $n_a\in\mathbb{Z}+\epsilon\, ,  \epsilon=0,{1\over 2}$,  and
$$
\prod_{a=1}^8 g_a=(pq)^2, \qquad \sum_{a=1}^8n_a=0
$$
(for notation, see \eqref{shorts}). Note that the variable $\epsilon$ cannot be removed from the definition of this function by shifts of other discrete parameters.

The basic symmetry transformation of the $V_\epsilon^{(r)}$-function is a direct generalization of
the relation \eqref{vt1}:
\begin{eqnarray}\label{E7trafo1} &&  \makebox[-1em]{}
V_\epsilon^{(r)}(\underline{g},\underline{n};p,q)=V_\delta^{(r)}(\underline{\tilde{g}},\underline{\tilde{n}} ;p,q)
\\ && \makebox[2em]{} \times
\prod_{1\leq b<c\leq 4}\Gamma^{(r)} (g_bg_c,n_b+n_c;p,q)
\Gamma^{(r)} (g_{b+4}g_{c+4},n_{b+4}+n_{c+4};p,q),
\nonumber\end{eqnarray}
where
$$
\left\{
\begin{array}{cl}
\tilde{g}_a =f g_a,&  a=1,2,3, 4,  \\
\tilde{g}_a = f^{-1} g_a, &  a=5,6,7, 8,
\end{array}
\right.;
\quad f=\sqrt{\frac{pq}{g_1g_2g_3g_4}}=\sqrt{\frac{g_5g_6g_7g_8}{pq}},
$$
\begin{equation}
\left\{
\begin{array}{cl}
\tilde{n}_a= n_a-\frac{1}{2}(\sum_{b=1}^4n_b), &a=1,2,3, 4,  \\
\tilde{n}_a= n_a+\frac{1}{2}(\sum_{b=1}^4n_b), &  a=5,6,7, 8,
\end{array}
\right.
\label{ntok}\end{equation}
and is it assumed that $|g_a|, |\tilde{g}_a|<1$. Here $\delta=0,{1\over 2}$, and one should take $\delta=\epsilon$, if $\sum_{b=1}^4n_b$  is an even integer, or otherwise $\delta\neq \epsilon$.

Define the following rarefied hyperbolic hypergeometric function:
\bea\label{wfunc}
W_{\epsilon}(\underline{s},\underline{n};\mathbf{\omega})=\int_{-\textup{i}\infty}^{\textup{i}\infty}\sum_{m\in \mathbb{Z}_r+\epsilon}{\prod_{a=1}^8\Lambda(s_a\pm y;n_a\pm m;\mathbf{\omega})
\over \Lambda(\pm 2y;\pm 2m;\mathbf{\omega})}{dy\over 2\textup{i}r\sqrt{\omega_1\omega_2}}\,,\quad
\eea
where $\Lambda(x\pm y;n\pm m;\mathbf{\omega})=\Lambda(x+ y;n+m;\mathbf{\omega})\Lambda(x-y;n-m;\mathbf{\omega})$
and the following balancing constraints on the parameters $s_j$ and $n_j$ hold true:
$$
\sum_{j=1}^8 s_j=2Q,\qquad \sum_{j=1}^8 n_j=0.
$$
Using the limit (\ref{limgg}), one can show that the transformation (\ref{E7trafo1}) reduces to
a symmetry relation for  function  (\ref{wfunc}) of the form:
\bea\label{parsym1} &&
W_{\epsilon}(\underline{s},\underline{n};\mathbf{\omega})=
W_{\delta}(\underline{\tilde{s}},\underline{\tilde{n}};\mathbf{\omega})
\\\nonumber &&  \makebox[1em]{}
\times \prod_{1\leq j< k \leq 4}\Lambda(s_j+s_k;n_j+n_k;\mathbf{\omega})
\prod_{5\leq j< k \leq 8}\Lambda(s_j+s_k;n_j+n_k;\mathbf{\omega})\, ,
\\ &&
\tilde{s}_j=s_j+\xi, \quad \tilde{s}_{j+4}=s_{j+4}-\xi, \quad j=1,2,3,4,\quad
\xi={1\over 2}(\omega_1+\omega_2-\sum_{j=1}^4 s_j)
\nonumber\eea
with the same $\tilde{n}_a$ as defined in (\ref{ntok}).
Now we parametrize the variables $n_a$ and $s_a$ in this identity in the following way:
\bea\label{limit11} &&
s_{1,2,5,6}=\gamma_{1,2,3,4}+\textup{i}\mu, \qquad  n_{1,2,5,6}=t_{1,2,3,4},
\\ \nonumber &&
s_{3,4,7,8}=\beta_{1,2,3,4}-\textup{i}\mu,\qquad n_{3,4,7,8}=l_{1,2,3,4},
\eea
where $\gamma_j$, $\beta_j$ and $l_j$, $t_j$ satisfy the constraints (\ref{mmm}).
Shifting the integration variable $y\to y-\textup{i}\mu$ on both sides and taking the limit $\mu\to -\infty$, we obtain from \eqref{parsym1}:
\bea \nonumber &&
J_{\epsilon}(\underline{\beta},\underline{l};\underline{\gamma},\underline{t})=e^{\pi\textup{i}
\left[\sum_{j=1}^4t_j-2(\epsilon+\delta)\right]}J_{\delta}(\underline{\tilde{\beta}},\underline{\tilde{l}};
\underline{\tilde{\gamma}},\underline{\tilde{t}})
\\ && \makebox[1em]{} \times
\prod_{j,k=1}^2\Lambda(\gamma_j+\beta_k;t_j+l_k;\mathbf{\omega})
\prod_{j,k=3}^4\Lambda(\gamma_j+\beta_k;t_j+l_k;\mathbf{\omega})\, ,
\label{jjsym1} \eea
where $J_{\epsilon}(\underline{\beta},\underline{l};\underline{\gamma},\underline{t})$ is defined in equation
(\ref{jparker}) and $J_{\delta}(\underline{\tilde{\beta}},\underline{\tilde{l}};\underline{\tilde{\gamma}},\underline{\tilde{t}})$
has the following arguments
\bea  \nonumber
&&\tilde{\beta}_a=\beta_a+\Theta(a)\eta\, , \quad \tilde{l}_a=l_a-\Theta(a)N,
\\ \nonumber
&&\tilde{\gamma}_a=\gamma_a+\Theta(a)\eta\, , \;\quad \tilde{t}_a=t_a-\Theta(a)N\, ,\quad a=1,2,3,4\, ,
\eea
where $\Theta(a)$  is the sign function taking the values
\be\label{thesig}
\Theta(a)=1\, , \quad a=1,2 \quad {\rm and} \quad \Theta(a)=-1\, , \quad a=3,4\,
\ee
and
$$
N={1\over 2}(t_1+t_2+l_1+l_2), \quad
\eta={1\over 2}(Q-\gamma_1-\gamma_2-\beta_1-\beta_2).
$$
As before, the discrete variable $\delta$ should be determined from the requirement for $N+\epsilon+\delta$
be integer (recall that  in \eqref{jjsym1} $t_a, l_a\in \mathbb{Z}+\epsilon$ and
$\tilde{t}_a, \tilde{l}_a\in \mathbb{Z}+\delta$).

In fact, one can eliminate $\epsilon$ and $\delta$ from \eqref{jjsym1} by shifting parameters $l_a$ and $t_a$,
as described after formula \eqref{jparker}, and shifting the summation variable $m\to m+N-\epsilon$ in $J_{\delta}(\underline{\tilde{\beta}},\underline{\tilde{l}};\underline{\tilde{\gamma}},\underline{\tilde{l}})$, which is allowed by the periodicity of the summand $m\to m+r$ in \eqref{jparker}.
This yields the identity
\bea\label{aza1}
&&J_0(\underline{\beta},\underline{l};\underline{\gamma},\underline{t})=e^{\textup{i}\pi
\left[t_3+t_4+l_1+l_2\right]}J_0(\underline{\tilde{\beta}},
\underline{l^{\ast}};\underline{\tilde{\gamma}},\underline{t^{\ast}})\nonumber
\\ \label{jjsym11} && \makebox[2em]{} \times
\prod_{j,k=1}^2\Lambda(\gamma_j+\beta_k;t_j+l_k;\mathbf{\omega})
\prod_{j,k=3}^4\Lambda(\gamma_j+\beta_k;t_j+l_k;\mathbf{\omega})\, ,
\\  \nonumber &&
l_a^{\ast}=l_a-(\Theta(a)+1)N, \qquad t_a^{\ast}=t_a-(\Theta(a)-1)N,\quad a=1,2,3,4,
\eea
which is a generalization of the relation \eqref{ide1b}
for the parafermionic hypergeometric integral \eqref{jparker}.

\subsection{Parafermionic symmetry ${\rm I}_b$}

Define a parafermionic analogue of the function \eqref{ehrho}:
\bea\nonumber
E_{\epsilon}(\underline{\rho},\underline{v};\mathbf{\omega})=\int_{-\textup{i}\infty}^{\textup{i}\infty}\sum_{m\in \mathbb{Z}_r+\epsilon}{\prod_{a=1}^6\Lambda(\rho_a\pm y;v_a\pm m;\mathbf{\omega})
\over \Lambda(\pm 2y;\pm 2m;\mathbf{\omega})}{dy\over 2\textup{i}r\sqrt{\omega_1\omega_2}},
\eea
where $v_a\in \mathbb{Z}+\epsilon$, $a=1,\ldots, 6$, and the shorthand is fixed in \eqref{shorts}.
In order to reduce relation (\ref{parsym1}), we use the parametrization
\bea &&\nonumber
s_{1,2,3,8}=\beta_{1,2,3,4}+\textup{i}\mu, \qquad n_{1,2,3,8}=l_{1,2,3,4},
\\  \nonumber  &&
s_{4,5,6,7}=\gamma_{4,1,2,3}-\textup{i}\mu,\qquad n_{4,5,6,7}=t_{4,1,2,3},
\eea
with the balancing condition (\ref{mmm}), shift the integration variable $y\to y+\textup{i}\mu$ only in the
left-hand side integral, and take the limit  $\mu\to \infty$. This procedure leads to the identity
\bea\label{azaz}
J_{\epsilon}(\underline{\beta}, \underline{l};\underline{\gamma},\underline{t})=
e^{\textup{i}\pi(t_4+l_4)}\prod_{i=1}^3\Lambda(\beta_i+\gamma_4;l_i+t_4;\mathbf{\omega})
\Lambda(\gamma_i+\beta_4;t_i+l_4;\mathbf{\omega})E_{\delta}(\underline{\rho},\underline{v};\mathbf{\omega}),
\eea
where $t_a, l_a\in \mathbb{Z}+\epsilon$, $v_a\in \mathbb{Z}+\delta$, and
\bea\nonumber
&&\rho_a=\beta_a+\xi\, ,\quad v_a=l_a-L\, ,\quad a=1,2,3\, ,\quad L={1\over 2}(l_1+l_2+l_3+t_4),
\\ \nonumber
&&\rho_a=\gamma_a-\xi\, ,\quad v_a=t_a+L\, ,\quad a=4,5,6\, ,\quad \xi={1\over 2}(\omega_1+\omega_2-\gamma_4-\sum_{j=1}^3 \beta_j).
\eea
The parameter $\delta$ again should be taken equal to $\epsilon$, if $L$ is an integer, and not equal to $\epsilon$, if $L$ is a half-integer. The identity \eqref{azaz} is a generalization of relation
\eqref{jheh} to the rarefied hyperbolic hypergeometric functions.

\subsection{Parafermionic symmetry II}

The second identity for the rarefied elliptic hypergeometric function is obtained from \eqref{E7trafo1}
by a group composition and it is a generalization of relation \eqref{vt2}:
\begin{eqnarray}\label{E7trafo2} && 
V_\epsilon^{(r)}(\underline{g},\underline{n};p,q)
= \prod_{1\leq b,c\leq 4}\Gamma^{(r)} (g_bg_{c+4},n_b+n_{c+4};p,q)\,
V_\rho^{(r)}\left(\underline{\hat{g}},\underline{\hat{n}};p,q\right),
\end{eqnarray}
where
$$
\hat{g}_a= \frac{\sqrt{ t_1t_2t_3t_4 }}{g_a}, \; a=1,2,3,4,  \quad
\hat{g}_a=  \frac{\sqrt{ t_5t_6t_7t_8 }}{g_a} , \; a=5,6,7,8,
$$
with $|g_a|, |\hat{g}_a|<1$ and
\be\label{newk}
\left\{
\begin{array}{cl}
\hat{n}_a= -n_a+\frac{1}{2}\left(\sum_{\ell=1}^4 n_\ell\right), & a=1,2,3,4,  \\
\hat{n}_a= -n_a-\frac{1}{2}\left(\sum_{\ell=1}^4 n_\ell\right), & a=5,6,7,8.
\end{array}
\right.
\ee
As before, the discrete parameters are chosen as $\rho=\epsilon$, if $\sum_{\ell=1}^4 n_\ell$  is
an even integer, or $\rho\neq\epsilon$ otherwise.

Using the limit (\ref{limgg}) one can show that relation (\ref{E7trafo2}) implies  that the function (\ref{wfunc}) satisfies the following second symmetry relation:
\bea\label{gg} &&
W_{\epsilon}(\underline{s},\underline{n};\mathbf{\omega})=W_{\rho}(\underline{\hat{s}},\underline{\hat{n}};\mathbf{\omega})
\prod_{1\leq b,c\leq 4}\Lambda(s_b+s_{c+4};n_b+n_{c+4};\mathbf{\omega}),
\\ && \makebox[2em]{}
\hat{s}_j=G-s_j, \quad \hat{s}_{j+4}=Q-G-s_{j+4}, \quad j=1,2,3,4,\quad
G=\frac{1}{2}\sum_{j=1}^4 s_j,
\nonumber\eea
and $\hat{n}_a$ are given in (\ref{newk}).
Applying to (\ref{gg}) the limiting procedure (\ref{limit11})  we obtain:
\bea \nonumber &&
J_{\epsilon}(\underline{\beta},\underline{l};\underline{\gamma},\underline{t})=e^{2\textup{i}\pi M}J_{\rho}(\underline{\hat{\beta}},\underline{\hat{l}};\underline{\hat{\gamma}},\underline{\hat{t}})
\\&& \makebox[2em]{} \times
\prod_{j,k=1}^2\Lambda(\gamma_j+\beta_{k+2};t_j+l_{k+2};\mathbf{\omega})
\Lambda(\gamma_{j+2}+\beta_k;t_{j+2}+l_k;\mathbf{\omega}),
\label{symII}\eea
where the arguments of function
$J_{\rho}(\underline{\hat{\beta}},\underline{\hat{l}};\underline{\hat{\gamma}},\underline{\hat{t}})$ can be written as
\bea \nonumber
&&\hat{\beta}_a=Q/2+\Theta(a)(G-Q/2)-\gamma_a\, , \quad \hat{l}_a=-t_a+\Theta(a)M, \quad
\\
&&\hat{\gamma}_a=Q/2+\Theta(a)(G-Q/2)-\beta_a\, , \;\quad \hat{t}_a=-l_a+\Theta(a)M,\quad a=1,2,3,4,
\label{shapki}\eea
with $\Theta(a)$  defined in \eqref{thesig}  and
\be\label{m1}
M={1\over 2}(t_1+t_2+l_1+l_2), \qquad
G={1\over 2}(\gamma_1+\gamma_2+\beta_1+\beta_2).
\ee
Again, here  $\rho$ is equal to $\epsilon$ if  $M$ is an integer, or $\rho\neq\epsilon$ otherwise.
Recall that in \eqref{symII} $t_a, l_a\in \mathbb{Z}+\epsilon$ and  $\hat{t}_a, \hat{l}_a\in \mathbb{Z}+\rho$.

As in the previous section, one can exclude from formula \eqref{symII}
both $\epsilon$ and $\delta$ parameters  by the shifts of parameters $t_a$,$l_a$, and of the summation variable,
as indicated after formula \eqref{jparker}, and obtain the equality
\bea \nonumber &&
J_0(\underline{\beta},\underline{l};\underline{\gamma},\underline{t})=e^{2\textup{i}\pi M}J_0(\underline{\hat{\beta}},\underline{l'};\underline{\hat{\gamma}},\underline{t'})
\\ \label{symIIc} && \makebox[2em]{}
\times\prod_{j,k=1}^2\Lambda(\gamma_j+\beta_{k+2};t_j+l_{k+2};\mathbf{\omega})
\Lambda(\gamma_{j+2}+\beta_k;t_{j+2}+l_k;\mathbf{\omega}),
\\ \label{shapkii} &&
\hat{\beta}_a=Q/2+\Theta(a)(G-Q/2)-\gamma_a, \qquad l_a'=-t_a+(\Theta(a)-1)M,
\\ \nonumber &&
\hat{\gamma}_a=Q/2+\Theta(a)(G-Q/2)-\beta_a, \qquad t_a'=-l_a+(\Theta(a)+1)M,\quad a=1,2,3,4.
\eea
Here $t_a, t_a', l_a, l_a' \in \mathbb{Z}$.

Let us rewrite  formula (\ref{symIIc}) in the parametrization (\ref{ptparam}).
Inserting $\gamma_a^{\circ}$ and $\beta_a^{\circ}$ , $a=1,2,3,4$,
defined in (\ref{ptparam}), we obtain:
\bea\label{fertv}
J_0(\underline{\beta}^{\circ},\underline{l};\underline{\gamma}^{\circ},\underline{t})=e^{2\textup{i}\pi M}
\Omega(\underline{\alpha},\underline{t},\underline{l})J_0(\underline{\beta}^{\diamond},\underline{l'},
\underline{\gamma}^{\diamond},\underline{t'}),
\eea
where $\beta_a^{\diamond}$ and $\gamma_a^{\diamond}$ are defined in \eqref{tvparam}, $l_a'$,  $t_a'$
are the same as in \eqref{shapkii}, and
\bea\nonumber &&
\Omega(\underline{\alpha},\underline{t},\underline{l})=
\Lambda(Q+\alpha_s-\alpha_3-\alpha_4;l_1+t_{3};\mathbf{\omega})
\Lambda(Q+\alpha_s-\alpha_1-\alpha_2;l_1+t_{4};\mathbf{\omega})
\\ \nonumber && \makebox[2em]{} \times
\Lambda(Q-\alpha_t+\alpha_2-\alpha_3;l_2+t_{3};\mathbf{\omega})
\Lambda(Q-\alpha_t+\alpha_4-\alpha_1;l_2+t_{4};\mathbf{\omega})
\\ \nonumber && \makebox[2em]{} \times
\Lambda(-Q+\alpha_t+\alpha_2+\alpha_3;l_3+t_{1};\mathbf{\omega})
\Lambda(Q-\alpha_s+\alpha_3-\alpha_4;l_4+t_{1};\mathbf{\omega})
\\ && \makebox[2em]{} \times
\Lambda(-Q+\alpha_t+\alpha_4+\alpha_1;l_3+t_{2};\mathbf{\omega})
\Lambda(Q-\alpha_s+\alpha_1-\alpha_2;l_4+t_{2};\mathbf{\omega}).
\label{umka} \eea
Formula (\ref{fertv}) is a parafermionic generalization of formula (\ref{tvfor}).

\subsection{Parafermionic symmetry III}

The third symmetry transformation for rarefied elliptic hypergeometric function
is obtained after equating the right-hand side
expressions in \eqref{E7trafo1} and \eqref{E7trafo2}:
\begin{eqnarray}\label{E7trafo3} && 
V_\epsilon^{(r)}(\underline{g},\underline{n};p,q)
= \prod_{1\leq b<c\leq 8}\Gamma^{(r)} (g_bg_c,n_b+n_c;p,q)\,
V_\epsilon^{(r)}\left(\frac{\sqrt{pq}}{\underline{g}},-\underline{n};p,q\right).
\end{eqnarray}
Applying the limit (\ref{limgg}) to (\ref{E7trafo3}) one can deduce for function (\ref{wfunc}) the third symmetry transformation
\begin{equation}\label{WIII}
W_{\epsilon}(\underline{s}, \underline{n};\mathbf{\omega})=W_{\epsilon}(Q/2-\underline{s},-\underline{n};\mathbf{\omega})
\prod_{1\leq j< k \leq 8}\Lambda(s_j+s_k;n_j+n_k;\mathbf{\omega}).
\ee
Parametrize $s_a$ and $n_a$ as:
\begin{equation}
s_j = \gamma_j+{\rm i}\mu, \qquad s_{j+4}=\beta_j-{\rm i}\mu, \quad n_j=t_j,\quad n_{j+4}=l_j
\qquad j=1,\ldots, 4.
\label{prs}\ee
Shifting now the integration variables $y\to y-\textup{i}\mu$ on both sides of \eqref{WIII}
and taking the limit $\mu\to -\infty$, we come to the symmetry transformation
\bea\label{azazo3}
&& J_{\epsilon}(\underline{\beta},\underline{l};\underline{\gamma};\underline{t})
=e^{\textup{i}\pi\left[\sum_{j=1}^4t_j\right]}
\prod_{j,k=1}^4\Lambda(\gamma_j+\beta_k;t_j+l_k;\mathbf{\omega})
\\ \nonumber
&&\times\int_{-\textup{i}\infty}^{\textup{i}\infty}\sum_{m\in \mathbb{Z}_r+\epsilon}
\prod_{a=1}^4\Lambda(\frac{Q}{2}-y-\gamma_a;-t_a-m;\mathbf{\omega})
\Lambda(\frac{Q}{2}+y-\beta_a;-l_a+m;\mathbf{\omega}){dy\over \textup{i}r\sqrt{\omega_1\omega_2}}.
\eea
This is a parafermionic extension of the standard hyperbolic identity \eqref{infy}.

\section{Parafermionic difference-recurrence equation}

As shown in \cite{Spiridonov:2016uae}, the function
$$
U_\epsilon(\underline{g},\underline{n}):=\frac{V_\epsilon^{(r)}(\underline{g},\underline{n})}
{\prod_{k=1}^2\Gamma^{(r)} (g_kg_3^{\pm 1},n_k\pm n_3;p,q)}
$$
satisfies the following mixed difference-recurrence equation:
\begin{eqnarray}\nonumber
&& \makebox[0em]{}
\mathcal{A}\left(\textstyle{\frac{g_1}{q^{n_1}},\frac{g_2}{q^{n_2}},
\ldots,\frac{g_8}{q^{n_8}},p;q^r}\right)
\Big(U_\epsilon(pg_1,p^{-1}g_2,n_1-1,n_2+1)-U_\epsilon(g_a,n_a)\Big)
\\ \nonumber && \makebox[1em]{}
+\mathcal{A}\left(\textstyle{\frac{g_2}{q^{n_2}},\frac{g_1}{q^{n_1}},
\ldots,\frac{g_8}{q^{n_8}},p;q^r}\right)
\Big(U_\epsilon(p^{-1}g_1,pg_2,n_1+1,n_2-1)-U_\epsilon(g_a,n_a)\Big)
\\ && \makebox[4em]{}
+ U_\epsilon(g_a,n_a)=0\, ,
\label{reheq}\end{eqnarray}
where the $\mathcal{A}$-potential has the form
$$
 \mathcal{A}(g_1,\ldots, g_8,p;q^r):=\frac{\theta\left(\frac{g_1}{pq^{1-r}g_3},
g_3g_1,\frac{g_3}{g_1};q^r\right)}
                 {\theta\left(\frac{g_1}{g_2},
\frac{g_2}{pq^{1-r}g_1},\frac{g_1g_2}{pq^{1-r}};q^r\right)}
\prod_{a=4}^8\frac{\theta\left(\frac{g_2g_a}{pq^{1-r}};q^r\right)}
{\theta\left(g_3g_a;q^r\right)}
$$
(for notation, see \eqref{shorts}).
Applying now to equation (\ref{reheq}) the limits (\ref{limgg}) and (\ref{limthet}), we obtain:
\be\label{urhyp}
{\cal B}(\underline{s},\underline{n};\omega_1,\omega_2)(Z(s_1+\omega_2,s_2-\omega_2,n_1-1,n_2+1)-Z(s))
+(s_1, n_1\leftrightarrow s_2,n_2)+Z(s)=0\, ,
\ee
where
\bea\nonumber &&
{\cal B}(\underline{s},\underline{n};\omega_1,\omega_2)={\sin{{\pi\over r\omega_1}}(s_1-s_3-\omega_2+(n_3-n_1+r-1)\omega_1)
\over \sin{{\pi\over r\omega_1}}(s_1-s_2+(n_2-n_1)\omega_1)}
\\ \nonumber && \makebox[-2em]{} \times
\frac{\sin{{\pi\over r\omega_1}}(s_1+s_3-(n_1+n_3)\omega_1)\sin{{\pi\over r\omega_1}}(s_3-s_1+(n_1-n_3)\omega_1)}
{\sin{{\pi\over r\omega_1}}(s_2-s_1-\omega_2+(n_1-n_2+r-1)\omega_1)\sin{{\pi\over r\omega_1}}(s_1+s_2-\omega_2-(n_1+n_2+1-r)\omega_1)}
\\ \nonumber && \makebox[2em]{} \times
\prod_{k=4}^8{\sin{{\pi\over r\omega_1}}(s_2+s_k-\omega_2-(n_2+n_k+1-r+\epsilon)\omega_1)\over \sin{{\pi\over r\omega_1}}(s_3+s_k-(n_3+n_k)\omega_1)},
\eea
$$
Z(\underline{s},\underline{n})={W_{\epsilon}(s_i,n_i)\over \Lambda(s_1\pm s_3,n_1\pm n_3)
\Lambda(s_2\pm s_3,n_2\pm n_3)}\, ,
$$
with the balancing condition:
$$
\sum_{i=1}^8 s_i=2(\omega_1+\omega_2),\,\quad \sum_{i=1}^8 n_i=0\, .
$$

Applying to (\ref{urhyp}) the limiting procedure (\ref{prs}), we obtain
\bea\nonumber &&
{\cal D}(\underline{\gamma},\underline{\beta},\underline{l},\underline{t})({\mathcal K}(\gamma_1+\omega_2,\gamma_2-\omega_2,t_1-1,t_2+1)-{\cal K}(\underline{\gamma},\underline{\beta},\underline{l},\underline{t})
\\ && \makebox[2em]{}
+(\gamma_1\leftrightarrow \gamma_2,t_1\leftrightarrow t_2)+{\cal K}(\underline{\gamma},\underline{\beta},\underline{l},\underline{t})=0,
\label{parvenu23}\eea
where
$$
{\cal K}(\underline{\gamma},\underline{\beta},\underline{l},\underline{t})
={J_{\epsilon}(\underline{\beta},\underline{l};\underline{\gamma};\underline{t})\over \Lambda(\gamma_1-\gamma_3,t_1-t_3)\Lambda (\gamma_2- \gamma_3,t_2-t_3 )},
$$
and
\bea\nonumber && \makebox[-1em]{}
{\cal D}(\underline{\gamma},\underline{\beta},\underline{l},\underline{t})={\sin{{\pi\over r\omega_1}}(\gamma_1-\gamma_3-\omega_2+(t_3-t_1+r-1)\omega_1)\sin{{\pi\over r\omega_1}}(\gamma_3-\gamma_1+(t_1-t_3)\omega_1)
\over \sin{{\pi\over r\omega_1}}(\gamma_1-\gamma_2+(t_2-t_1)\omega_1)\sin{{\pi\over r\omega_1}}(\gamma_2-\gamma_1-\omega_2+(t_1-t_2+r-1)\omega_1)}\quad\quad\\ \nonumber
&& \makebox[6em]{}  \times
\prod_{k=1}^4{\sin{{\pi\over r\omega_1}}(\gamma_2+\beta_k-\omega_2-(t_2+l_k+1-r+\epsilon)\omega_1)\over \sin{{\pi\over r\omega_1}}(\gamma_3+\beta_k-(t_3+l_k+\epsilon)\omega_1)}.
\eea
Equation \eqref{parvenu23} is a parafermionic extension of the difference equation \eqref{secdif}.

\section{Supersymmetric Racah-Wigner symbols}

In this section we specialize some of the above results to the case $r=2$.
We show that the $r=2$ rarefied hyperbolic gamma function  $\Lambda(y,m;\mathbf{\omega})$ is
a building block for the fusion matrix of the $N=1$ supersymmetric Liouville theory
and, closely related to it, Racah-Wigner symbols of the quantum supergroup ${\rm U}_q({\rm osp}(1|2))$
for the continuous series representations.
The fusion matrix  of the Neveu-Schwarz  sector of $N=1$ supersymmetric Liouville theory was studied in
\cite{Hadasz:2007wi,Chorazkiewicz:2008es,Hadasz:2013bwa}.
There were suggested some integrals of linear combinations of the product of eight, so-called,
Neveu-Schwarz and Ramond hyperbolic gamma functions, $S_{\rm NS}(x)$ and $S_{\rm R}(x)$,
in the parametrization \eqref{ptparam}. First, we review some basics of the $N=1$ super Liouville theory.
Then describe relations of $S_{\rm NS}(x)$ and $S_{\rm R}(x)$ to
$\Lambda(y,m;\mathbf{\omega})$ for $r=2$  and show that all the suggested integrals coincide with the
parafermionic hypergeometric function \eqref{jparker} for $r=2$ with the corresponding match of parameters.
In \cite{Pawelkiewicz:2013wga}, an expression for the universal supersymmetric Racah-Wigner symbols
was suggested in the parametrization \eqref{tvparam}. Using formula \eqref{fertv} we show that the
restriction of these symbols to the Neveu-Schwarz  sector is in agreement
with the expression proposed in  \cite{Hadasz:2013bwa} and also find its general expression
in parametrization \eqref{ptparam}.

\subsection{Some basics on $N=1$ super Lioville field theory}

$N=1$ super Liouville field theory is defined on a two-dimensional surface with
the metric $g_{ab}$ by the local Lagrangian density
\be
{\cal L}={1\over 2\pi}g_{ab}\partial_a\varphi\partial_b \varphi+
{1\over 2\pi}(\psi\bar{\partial}\psi+\bar{\psi}\partial\bar{\psi})+2\textup{i}\mu b^2\bar{\psi}\psi e^{b\varphi}+
2\pi \mu^2 b^2 e^{2b\varphi}.
\ee
The energy-momentum tensor and the superconformal current are
\bea
T=-{1\over 2} (\partial\varphi\partial\varphi-Q\partial^2\varphi+\psi\partial\psi),
\quad G=\textup{i}(\psi\partial\varphi-Q\partial\psi),
\eea
where $Q=b+b^{-1}$. The modes $L_n=\oint {dz\over 2\pi \textup{i}}z^{n+1}T(z)$ and $G_k
=\oint {dz\over 2\pi \textup{i}}z^{k+1/2}G(z)$ satisfy the superconformal algebra
 \bea \nonumber
&& [L_m,L_n]=(m-n)L_{m+n}+{c\over 12}m(m^2-1)\delta_{m+n},
\\  &&
[L_m, G_k]={m-2k\over 2}G_{m+k},
\\   &&
\{G_k,G_l\}=2L_{l+k}+{c\over 3}\left(k^2-{1\over 4}\right)\delta_{k+l},
  \nonumber \eea
with the central charge $c=c_{SL}={3\over 2}+3Q^2.$
Here $k$ and $l$ take integer values for the Ramond algebra and half-integer values for the Neveu-Schwarz algebra.

Since in the Neveu-Schwarz sector we have supercurrent generators $G_k$ with half-integer $k$,
descendant fields are broken into two representations of the plain Virasoro algebra
of integer and half-integer level. Thus in the Neveu-Schwarz sector there are two types of Virasoro
primary fields.  The first type is associated with the vertex operator
$N_{\alpha}=e^{\alpha \varphi}$ and has the conformal dimension
$ \Delta_{\alpha}={1\over 2}\alpha(Q-\alpha).$
The physical states have $\alpha={Q\over 2}+\textup{i}P$.

The second type primary field is given by the supercurrent descendant
$ \tilde{N}_{\alpha}=G_{-1/2}N_{\alpha} $
and has the conformal dimension
$$
\Delta_{\tilde{\alpha}}={1\over 2}\alpha(Q-\alpha)+{1\over 2}.
$$

The Ramond vertex operators and their dimensions are
$$
R_{\alpha}^{\pm}=\sigma^{\pm} e^{\alpha \varphi},
\quad \Delta_{\alpha}^{R^{\pm}}={1\over 16}+{1\over 2}\alpha(Q-\alpha),
$$
where $\sigma^{\pm}$ are the spin fields satisfying the property:
$$
\psi(z)\sigma^{\pm}(0)\sim{\sigma^{\mp}(0)\over \sqrt{z}}
$$
The Neveu-Schwarz and Ramond operators with the same conformal dimensions are proportional to each other,
namely, one has
$ N_{\alpha}\propto N_{Q-\alpha},\, R_{\alpha}\propto R_{Q-\alpha}. $

In the paper \cite{Hadasz:2013bwa}, $6j$-symbols of continuous series
representation of the quantum supergroup ${\rm U}_q({\rm osp}(1|2))$ were denoted as
$\big\{\,{}^{\alpha_1}_{\alpha_3}\,{}^{\alpha_2}_{{\alpha_4}}\,\big|\,
{}^{\alpha_s}_{\alpha_t}\big\}_{\nu_1\nu_2}^{\nu_3\nu_4}$
(we drop the subindex $b$ indicating that we deal with the quantum group).
Here $\alpha_j$, $j=1,\ldots,4$, $\alpha_s$, $\alpha_t$ are the continuous spins fixing
corresponding representations, and $\nu_k=0,1$, $k=1,\ldots,4,$
take into account a doubling of the Clebsch-Gordan coefficients following from the existence
of two different intertwining operators for the decomposition of tensor products of two representations with fixed $\alpha_1$ and $\alpha_2$.
The following dictionary is established between the $6j$-symbols of
${\rm U}_q({\rm osp}(1|2))$ and the fusion matrix of $N=1$ supersymmetric Liouville field
theory in the Neveu-Schwarz sector  \cite{Hadasz:2013bwa}:
\bea \nonumber &&
 F_{N_{\alpha_s},N_{\alpha_t}}
 \Big[{}^{N_{\alpha_3}}_{N_{\alpha_4}}\, {}^{N_{\alpha_2}}_{N_{\alpha_1}} \Big]
\propto\big\{\,{}^{\alpha_1}_{\alpha_3}\,{}^{\alpha_2}_{{\alpha_4}}\,\big|\,{}^{\alpha_s}_{\alpha_t}\big\}_{11}^{11},
\quad
 F_{N_{\alpha_s},\tilde{N}_{\alpha_t}}
 \Big[{}^{N_{\alpha_3}}_{N_{\alpha_4}}\, {}^{N_{\alpha_2}}_{N_{\alpha_1}} \Big]
\propto\big\{\,{}^{\alpha_1}_{\alpha_3}\,{}^{\alpha_2}_{{\alpha_4}}\,\big|\,{}^{\alpha_s}_{\alpha_t}\big\}_{00}^{11},
\\ &&
 F_{\tilde{N}_{\alpha_s},N_{\alpha_t}}
 \Big[{}^{N_{\alpha_3}}_{N_{\alpha_4}}\, {}^{N_{\alpha_2}}_{N_{\alpha_1}} \Big]
\propto\big\{\,{}^{\alpha_1}_{\alpha_3}\,{}^{\alpha_2}_{{\alpha_4}}\,\big|\,{}^{\alpha_s}_{\alpha_t}\big\}_{11}^{00},
\quad
F_{\tilde{N}_{\alpha_s},\tilde{N}_{\alpha_t}}\Big[
{}^{N_{\alpha_3}}_{N_{\alpha_4}}\, {}^{N_{\alpha_2}}_{N_{\alpha_1}}
\Big]\propto\big\{\,{}^{\alpha_1}_{\alpha_3}\,{}^{\alpha_2}_{{\alpha_4}}\,\big|\,
{}^{\alpha_s}_{\alpha_t}\big\}_{00}^{00}.
\eea
This dictionary shows that if one considers $6j$-symbols as fusion matrices, then $\alpha_j$ acquire
the meaning of momenta and $\nu_j$ distinguish between $N_{\alpha_s}$ and $\tilde{N}_{\alpha}$
primary field entries.

The general supersymmetric Racah-Wigner symbols were introduced in \cite{Pawelkiewicz:2013wga}. They were
denoted as  $\quad$ $\left\{ \begin{array}{cc  c}
  \alpha_1^{a_1} & \alpha_3^{a_3}  & \alpha_s^{a_s} \\
  \alpha_2^{a_2} & \alpha_4^{a_4}  & \alpha_t^{a_t}
  \end{array}  \right\}^{\nu_3 \nu_4}_{\nu_1 \nu_2}$,
where $\alpha^a$ does not mean a power of $\alpha$, but denotes a pair of variables---continuous $\alpha$
and discrete $a$. The choice $a_j=0$ corresponds to  the Neveu-Schwarz operator,
and $a_j=1$ is connected to the Ramond operator. For the Ramond operator $\nu_k=0,1$ distinguish between $R^{+}$ and $R^{-}$ operators. When all $a_j=0$, one gets the $6j$-symbols of ${\rm U}_q({\rm osp}(1|2))$.

The general parafermionic hypergeometric function \eqref{jparker} is expected to be a key ingredient of
the fusion matrix of the parafermionic Liouville field theory discussed in \cite{Bershtein:2010wz} with the
central charge $c={3r\over r+2}+{6\over r}(b+b^{-1})^2$. For $r=2$, this is indeed so for
the $N=1$ supersymmetric Liouville field theory, as follows from the comparison with the results of
 \cite{Hadasz:2013bwa}.

\subsection{Supersymmetric hypergeometric function and $6j$-symbols}

We start by describing the relation between $S_{\rm NS}(x)$, $S_{\rm R}(x)$ and
the function $\Lambda(y,m;\mathbf{\omega})$ for $r=2$ \cite{SS}.
Setting $\omega_2=b$ and $\omega_1=b^{-1}$, $Q=b+b^{-1}$,
and using the notation accepted in conformal field theory literature
$\gamma^{(2)}(z;b,1/b)=:S_b(z),$ we obtain
\bea\label{l1s}  &&
\Lambda(y,0;b^{-1},b)=S_b\left({y\over 2}\right)
S_b\left({y\over 2}+{Q\over 2}\right)\equiv S_{\rm NS}(y)\equiv S_{1}(y),
\\ && \label{l2s}
\Lambda(y,1;b^{-1},b)=S_b\left({y\over 2}+{b\over 2}\right)
S_b\left({y\over 2}+{b^{-1}\over 2}\right)\equiv S_{\rm R}(y)\equiv S_{0}(y).
\eea
The subscript $a$ of $S_a(y)$ is defined modulo 2: $S_{a+2}(y)\equiv S_a(y)$.
Using formula (\ref{lambdasign}), it is easy to see the relation between  $\Lambda(y,m;b^{-1},b)$
for $r=2$ and $S_a(y)$ for arbitrary $m$:
\be\label{lambdasign2}
\Lambda(y,m+2k;\mathbf{\omega})=(-1)^{mk}\Lambda(y,m;\mathbf{\omega})\, .
\ee
Formulae (\ref{l1s}), (\ref{l2s}), and (\ref{lambdasign2}) imply
$$
\Lambda(y,K;b^{-1},b)=
 S_{\rm NS}(y)\equiv S_{1}(y),\quad {\rm if}\quad  K \quad {\rm is}\quad {\rm even},
$$
$$
\Lambda(y,K;b^{-1},b)=(-1)^{K-1\over 2}
S_{\rm R}(y)\equiv S_{0}(y),\quad {\rm if}\quad  K \quad {\rm is}\quad {\rm odd},
$$
or, combining these equalities, we have
\be\label{l2s22}
\Lambda(y,K;b^{-1},b)=(-1)^{F(K)} S_{K+1}(y),\qquad F(K)={1-(-1)^K\over 2}{K-1\over 2}.
\ee

Now we are ready to discuss supersymmetric Racah-Wigner symbols for the supergroup
 ${\rm U}_q({\rm osp}(1|2))$ which have the following expression \cite{Hadasz:2013bwa}:
\be\label{ponsot}
\big\{\,{}^{\alpha_1}_{\alpha_3}\,{}^{\alpha_2}_{{\overline{\alpha}_4}}\,\big|\,{}^{\alpha_s}_{\alpha_t}\big\}_{\nu_1\nu_2}^{\nu_3\nu_4}=
{S_{\nu_4}(\alpha_s+\alpha_2-\alpha_1)S_{\nu_1}(\alpha_1+\alpha_t-\alpha_4)\over
S_{\nu_2}(\alpha_t+\alpha_2-\alpha_3)S_{\nu_3}(\alpha_3+\alpha_s-\alpha_4)}
I_{\alpha_s,\alpha_t}\left[\begin{array}{cc}
\alpha_3& \alpha_2\\
\alpha_4& \alpha_1 \end{array}\right]_{\nu_1\nu_2}^{\nu_3\nu_4},
\ee
where ${\overline{\alpha}_4}=Q-\alpha_4$, $\nu_j=0,1$, $j=1,2,3,4,$ $\sum_{i=1}^4\nu_i=0\, \, {\rm mod}\: 2$, and
\bea \nonumber && \makebox[-1em]{}
I_{\alpha_s,\alpha_t}\left[\begin{array}{cc}
\alpha_3& \alpha_2\\
\alpha_4& \alpha_1 \end{array}\right]_{\nu_1\nu_2}^{\nu_3\nu_4}=(-1)^{\nu_3\nu_2+\nu_4}\int_{-\textup{i}\infty}^{\textup{i}\infty}\sum_{\nu=0}^{1}
(-1)^{\nu(\nu_2+\nu_4)}
S_{1+\nu+\nu_3}(y+\gamma_1^{\circ})S_{1+\nu+\nu_4}(y+\gamma_2^{\circ})
\\ \nonumber && \makebox[2em]{} \times S_{1+\nu+\nu_3}(y+\gamma_3^{\circ})S_{1+\nu+\nu_4}(y+\gamma_4^{\circ})
S_{\nu}(-y+\beta_1^{\circ})S_{\nu+\nu_2+\nu_3}(-y+\beta_2^{\circ})
\\ && \makebox[2em]{} \times
S_{\nu+\nu_2+\nu_3}(-y+\beta_3^{\circ})
S_{\nu}(-y+\beta_4^{\circ})
{dy\over 2\textup{i}}.
\label{rwhps}\eea
We recall that parameters $\beta^{\circ}_a$ and $\gamma^{\circ}_a$ are defined in \eqref{ptparam}.
In fact, in order to rewrite $6j$-symbols expression of \cite{Hadasz:2013bwa} in the form (\ref{rwhps}),
we have replaced $\alpha_4$ by its reflected value, $\alpha_4\to {\overline{\alpha}_4}$
(which corresponds to the equivalent representation),
and we also shifted the integration variable by $-Q/2-\alpha_s$.

One can check that the integral (\ref{rwhps}) is a special case of the supersymmetric hypergeometric function (\ref{jparker}) for $r=2$ for an appropriate parametrization. Without loss of the generality, we take
 $\epsilon=0$ and set
\be
\begin{aligned}
&t_1=\nu_3,\\
&t_2=\nu_4,
\end{aligned}\qquad
\begin{aligned}
& t_3=\nu_3,\\
&t_4=-\nu_4,
\end{aligned}\qquad
\begin{aligned}
&l_1=-1,\\
&l_2=1+\nu_2-\nu_3,
\end{aligned}\qquad
\begin{aligned}
&l_3=1-\nu_2-\nu_3,\\
&l_4=-1,
\end{aligned}
\nonumber\ee
so that the condition $\sum_{i=1}^4(t_i+l_i)=0$ is satisfied. Originally, there were 6 independent
discrete variables among $t_j$ and $l_j$, and only three variables among $\nu_j$ are independent,
i.e. we have given to three discrete variables some fixed values. One thus obtains
\bea\nonumber &&
J_0(\beta_1^{\circ},-1,\beta_2^{\circ},1+\nu_2-\nu_3,\beta_3^{\circ},1-\nu_2-\nu_3,\beta_4^{\circ},-1;
\gamma_1^{\circ},\nu_3,\gamma_2^{\circ},\nu_4,\gamma_3^{\circ},\nu_3,\gamma_4^{\circ},\nu_4)
\\ \nonumber && \makebox[1em]{}
=\int_{-\textup{i}\infty}^{\textup{i}\infty}\sum_{\nu=0}^{1}
\Lambda(y+\gamma_1^{\circ},\nu_3+\nu)\Lambda(y+\gamma_2^{\circ},\nu+\nu_4)
\Lambda(y+\gamma_3^{\circ},\nu+\nu_3)
\\ \nonumber &&  \makebox[2em]{} \times
\Lambda(y+\gamma_4^{\circ},\nu-\nu_4)
\Lambda(-y+\beta_1^{\circ},-\nu-1)\Lambda(-y+\beta_2^{\circ}, -\nu+1+\nu_2-\nu_3)
\\  &&  \makebox[2em]{} \times
\Lambda(-y+\beta_3^{\circ},-\nu+1-\nu_2-\nu_3)
\Lambda(-y+\beta_4^{\circ},-\nu-1){dy\over 2\textup{i}},
\label{jrw} \eea
where we drop the dependence on $b$ in the notation. Expression (\ref{jrw}) coincides with (\ref{rwhps}).
To verify that, we should check only the sign factor, the rest is obvious. It is clear that the pair of the terms with the same spin structure will not produce any sign, so we can expect some sign factor only from the products
\be\label{paio1}
\Lambda(y+\gamma_2^{\circ},\nu+\nu_4)\Lambda(y+\gamma_4^{\circ},\nu-\nu_4)
\ee
and
\be\label{paio2}
\Lambda(-y+\beta_2^{\circ}, -\nu+1+\nu_2-\nu_3)\Lambda(-y+\beta_3^{\circ},-\nu+1-\nu_2-\nu_3)\, .
\ee
Consider the first product. It is clear that it can produce the minus sign only when $\nu_4=1$.
Setting $\nu_4=1$, we easily see from (\ref{l2s22}) that the minus sign appears only when $\nu=0$.
Collecting all together, we obtain:
\bea\nonumber
&&\Lambda(y+\gamma_2^{\circ},\nu+\nu_4)\Lambda(y+\gamma_4^{\circ},\nu-\nu_4)
=(-1)^{\nu_4(\nu+1)}S_{1+\nu+\nu_4}(y+\gamma_2^{\circ})S_{1+\nu+\nu_4}(y+\gamma_4^{\circ})\, .
 \nonumber\eea
Similarly, it is obvious that the product (\ref{paio2}) yields negative sign only if $\nu_2=1$.
Putting in (\ref{paio2}) $\nu_2=1$, one can check that the minus sign appears only when $\nu+\nu_3=1$.
Therefore we can write
\bea\nonumber
&&\Lambda(-y+\beta_2^{\circ}, -\nu+1+\nu_2-\nu_3)\Lambda(-y+\beta_3^{\circ},-\nu+1-\nu_2-\nu_3)
\\ &&  \makebox[2em]{}
=(-1)^{\nu_2(\nu+\nu_3)}S_{\nu+\nu_2+\nu_3}(-y+\beta_2^{\circ})S_{\nu+\nu_2+\nu_3}(-y+\beta_3^{\circ})\, .
\nonumber\eea
Noting that $(-1)^{\nu_4(\nu+1)}(-1)^{\nu_2(\nu+\nu_3)}=(-1)^{\nu_3\nu_2+\nu_4}(-1)^{\nu(\nu_2+\nu_4)}$,
we see that all the awkward sign factors in  (\ref{rwhps}) are coming from the sign
differences in $\Lambda(y,M)$ and $S_\nu(y)$ functions (\ref{l2s22}). This fact and similar ones described
below show that the use of our $\Lambda$-function is more convenient.

Now we employ relation \eqref{fertv} for function \eqref{jrw} to give another form of the integral \eqref{rwhps}.
From definition (\ref{m1}), we find $M=\tfrac12(\nu_2+\nu_4)$ and all spin structures in \eqref{shapkii}:
$$
\begin{aligned}
&l_1'=- t_1=-\nu_3,\\
&l_2'= -t_2=-\nu_4,
\end{aligned}\qquad
\begin{aligned}
&l_3'= -t_3-2M=-\nu_3-\nu_2-\nu_4,\\
&l_4'=-t_4-2M=-\nu_2,
\end{aligned}
$$
$$
\begin{aligned}
&t_1'=-l_1+2M=1+\nu_4+\nu_2,\\
&t_2'=-l_2+2M=-1+\nu_3+\nu_4,
\end{aligned}\qquad
\begin{aligned}
&t_3'=-l_3=-1+\nu_3+\nu_2,\\
&t_4'=-l_4=1.
\end{aligned}
$$
Shifting there additionally the summation variable $m$ by $-\nu_2$ and denoting it as $\nu$, we obtain for the integral standing on the right-hand
side of relation \eqref{fertv}:
\bea\nonumber && \makebox[-2em]{}
\tilde J_0(\underline{\alpha},\underline{\nu})\equiv
J_0(\underline{\beta}^{\diamond},\underline{l'},
\underline{\gamma}^{\diamond},\underline{t'})
=\int_{-\textup{i}\infty}^{\textup{i}\infty}\sum_{\nu=0}^{1}
\Lambda(-y-\alpha_{23t}, \nu_2-\nu_3-\nu)\Lambda(-y-\alpha_{14t}, \nu_2-\nu_4-\nu)
\\ \nonumber && \makebox[2em]{} \times
\Lambda(-y-\alpha_{12s}, -\nu_3-\nu_4-\nu)
\Lambda(-y-\alpha_{34s},-\nu)
\\ \nonumber && \makebox[2em]{} \times\Lambda(y+\alpha_{1234}, \nu_4+\nu+1)
\Lambda(y+\alpha_{13st}, -\nu_2+\nu_3+\nu_4+\nu-1)
\\  && \makebox[2em]{} \times
\Lambda(y+2Q, \nu_3+\nu-1)\Lambda(y+\alpha_{24st}, -\nu_2+\nu+1)
{dy\over 2\textup{i}}.
\label{bvbv3}
\eea
Using formulae from Appendix A for  $a_j=0$, we come to the expression:
\bea\nonumber && \makebox[-2em]{}
\tilde J_0(\underline{\alpha},\underline{\nu})=(-1)^{D}
\int_{-\textup{i}\infty}^{\textup{i}\infty}\sum_{\nu=0}^{1}
S_{\nu_2-\nu_3-\nu+1}(-y-\alpha_{23t})S_{\nu_2-\nu_4-\nu+1}(-y-\alpha_{14t})
\\ \nonumber && \makebox[2em]{} \times
S_{-\nu_4-\nu_3-\nu+1}(-y-\alpha_{12s})S_{-\nu+1}(-y-\alpha_{34s})S_{\nu_4+\nu}(y+\alpha_{1234})
\\ && \makebox[2em]{} \times
S_{-\nu_2+\nu_3+\nu_4+\nu}(y+\alpha_{13st})
S_{\nu_3+\nu}(y+2Q)S_{-\nu_2+\nu}(y+\alpha_{24st}) {dy\over 2\textup{i}},
\label{inttv85}\eea
where $D=\nu_2\nu_3\nu_4+\nu_2\nu_3+\nu_2\nu_4+\nu_3\nu_4+\nu_2+\nu_4.$
Compute also the prefactor $\Omega$ (\ref{umka}):
\bea
&&\Omega(\underline{\alpha},\underline{\nu})=(-1)^{\nu_3(\nu_2+\nu_4)}
S_{\nu_3}(Q+\alpha_s-\alpha_3-\alpha_4)S_{\nu_4}(Q+\alpha_s-\alpha_1-\alpha_2)
\\ \nonumber && \makebox[2em]{} \times
S_{\nu_2}(Q-\alpha_t+\alpha_2-\alpha_3)
S_{\nu_1}(Q-\alpha_t+\alpha_4-\alpha_1)S_{\nu_2}(-Q+\alpha_t+\alpha_2+\alpha_3)
\\ \nonumber && \makebox[2em]{}\times
S_{\nu_3}(Q-\alpha_s+\alpha_3-\alpha_4)
 S_{\nu_1}(-Q+\alpha_t+\alpha_4+\alpha_1)S_{\nu_4}(Q-\alpha_s+\alpha_1-\alpha_2),
\eea
so that one can write
\be\label{nutv}
I_{\alpha_s,\alpha_t}\left[\begin{array}{cc}
\alpha_3& \alpha_2\\
\alpha_4& \alpha_1 \end{array}\right]_{\nu_1\nu_2}^{\nu_3\nu_4}=(-1)^{\nu_2+\nu_4}
\Omega(\underline{\alpha},\underline{\nu})\tilde J_0(\underline{\alpha},\underline{\nu}),
\ee
where $\tilde J_0$-function is fixed in \eqref{inttv85}.

Now we can compare our results with the formula suggested  in \cite{Pawelkiewicz:2013wga} for the
universal supersymmetric Racah-Wigner symbols in the parametrization \eqref{tvparam}:
\begin{eqnarray}\nonumber
 \left\{ \begin{array}{cc  c}
  \alpha_1^{a_1} & \alpha_3^{a_3}  & \alpha_s^{a_s} \\
  \alpha_2^{a_2} & \alpha_4^{a_4}  & \alpha_t^{a_t}
  \end{array}  \right\}^{\nu_3 \nu_4}_{\nu_1 \nu_2}
 \hspace{-10pt} & =& \hspace{-1pt}
\, \mathcal{P}(\alpha_j, \nu_j)
\int_{\mathcal{C}} du \,
\sum_{\nu=0}^1
\Big(
  (-1)^{X} \,  S_{1+\nu_3 + \nu_4 + \nu}(u-\alpha_{12s})
\\[1mm]  \label{RW_s_a2}
&&  \hspace{-80pt}
\times  S_{1+ \nu }(u-\alpha_{s34})
S_{1+\nu_1+\nu_4+ \nu }(u-\alpha_{23t})
 S_{1+\nu_2+\nu_4+ \nu }(u-\alpha_{1t4})  \,
 \\[2mm]
  \nonumber
&&  \hspace{-80pt}
\times S_{\nu_4+ \nu }(\alpha_{1234} -u )S_{\nu_1 + \nu+ a_1}(\alpha_{st13} -u)
S_{\nu_2 + \nu + a_2 }(\alpha_{st24} -u) \,  S_{\nu_3 + \nu + a_s}(2Q - u) \Big)\, ,
\end{eqnarray}
where
\begin{equation} \label{signX}
(-1)^X =
 (-1)^{\nu( a_s \nu_1 + a_1 \nu_3 + a_4 \nu_4 +  a_1 a_s + a_2 a_4 + a_s+a_t )}\, ,
 \end{equation}
and the following selection rules hold true
\be\label{delnua}
\sum_{j=1}^4 \nu_j =a_s+a_t  \, (\rm{mod}\, 2),
\ee
\begin{equation}\label{aaa1}
a_s = a_1+ a_2 = a_3 + a_4 \, ({\rm mod}\, 2), \qquad
a_t = a_1 + a_4 = a_2 + a_3 \, ({\rm mod}\, 2).
\end{equation}
The prefactor $\mathcal{P}(\alpha_j, \nu_j)$ is a normalization of $6j$-symbols, it is represented by a complicated combination of square roots of various $S_{\nu}$-functions and is given explicitly in \cite{Pawelkiewicz:2013wga}.  For brevity we will not discuss it here. Parameters $a_i=0,1$ distinguish between the NS and the Ramond sectors: $a_i=0$ corresponds to primary fields
of the  NS-sector, and $a_i=1$ to those of the R-sector.
Expression \eqref{RW_s_a2} was suggested without proof, but it was tested using the limit when it reduces to
$6j$-symbols of the finite-dimensional representations of ${\rm U}_q({\rm osp}(1|2))$.

As clear from the definition, in the case, when all $a_j=0$, i.e. when all primaries lie in the NS sector and  (\ref{RW_s_a2}) gives $6j$-symbols of ${\rm U}_q({\rm osp}(1|2))$, this expression  should match with (\ref{ponsot}).
This was not proved in \cite{Pawelkiewicz:2013wga}, and now we can do it.

First, let us resolve the constraint  (\ref{delnua}) for $\nu_1$:
\be\label{nu1u}
\nu_1=a_s+a_t -\nu_2-\nu_3-\nu_4\, (\rm{mod}\, 2),
\ee
and insert in the integrand of  (\ref{RW_s_a2}), which yields
\begin{eqnarray}   \nonumber
 \left\{ \begin{array}{cc  c}
  \alpha_1^{a_1} & \alpha_3^{a_3}  & \alpha_s^{a_s} \\
  \alpha_2^{a_2} & \alpha_4^{a_4}  & \alpha_t^{a_t}
  \end{array}  \right\}^{\nu_3 \nu_4}_{\nu_1 \nu_2}
 \hspace{-10pt} & =& \hspace{-1pt}
\, \mathcal{P}(\alpha_i, \nu_i)
\int_{\mathcal{C}} du \,
\sum_{\nu=0}^1
\Big(
  (-1)^{X} \,  S_{1+\nu_3 + \nu_4 + \nu}(u-\alpha_{12s})
\\ \nonumber
&&  \hspace{-80pt}
\times  S_{1+ \nu }(u-\alpha_{s34})
S_{1+\nu_2+\nu_3-a_s-a_t+ \nu}(u-\alpha_{23t})
 S_{1+\nu_2+\nu_4+ \nu }(u-\alpha_{1t4})  \,
S_{\nu_4+ \nu}(\alpha_{1234} -u ) \\
&&  \hspace{-80pt}
\times S_{\nu_2+\nu_3+\nu_4 + \nu + a_2+a_t}(\alpha_{st13} -u)
S_{\nu_2 + \nu + a_2 }(\alpha_{st24} -u) \,  S_{\nu_3 + \nu + a_s}(2Q - u) \Big)\ .
\label{RWdlo}\end{eqnarray}
Here we used also the constraints (\ref{aaa1}). Setting all $a_j=0$
and identifying $u=-y$, one can see that integral in the right-hand side of \eqref{RWdlo} coincides with (\ref{inttv85}) up to the $(-1)^D$ sign factor.

Now we would like to show that the function standing in  (\ref{RW_s_a2}) after the $\mathcal{P}(\alpha_j, \nu_j)$ factor coincides with the special $r=2$ case of the parafermionic hypergeometric integral (\ref{jparker}).
Formula (\ref{bvbv3}) suggests the following generalization of the function
$\tilde J_0(\underline{\alpha},\underline{\nu})$ including the variables $a_j$:
\bea \nonumber && \makebox[-2em]{}
\tilde J_0(\underline{\alpha},\underline{\nu},\underline{a})
=\int_{-\textup{i}\infty}^{\textup{i}\infty}\sum_{\nu=0}^{1}
\Lambda(-y-\alpha_{23t}, \nu_2-\nu_3-\nu-a_s-a_t)\Lambda(-y-\alpha_{34s},-\nu)
\\ \nonumber &&  \makebox[1em]{}\times
\Lambda(-y-\alpha_{14t}, \nu_2-\nu_4-\nu)\Lambda(-y-\alpha_{12s}, -\nu_3-\nu_4-\nu)
\\ \nonumber && \makebox[1em]{} \times
\Lambda(y+\alpha_{1234}, \nu_4+\nu+1)
\Lambda(y+\alpha_{13st}, -\nu_2+\nu_3+\nu_4+\nu-1+a_2+a_t)
\\&&  \makebox[1em]{}\times
\Lambda(y+2Q, \nu_3+\nu-1+a_s)\Lambda(y+\alpha_{24st}, -\nu_2+\nu+1-a_2)
{dy\over 2\textup{i}}\, .
\label{generlam} \eea
Clearly the spin structures of \eqref{RWdlo} and \eqref{generlam} match and we should check
only the sign appearing in passage from $\Lambda(x,K)$ to $S_{K+1}(x)$.

Using formulae from appendix A, we can write
\begin{eqnarray} \nonumber &&
\tilde J_0(\underline{\alpha},\underline{\nu},\underline{a})
=(-1)^B\int_{-\textup{i}\infty}^{\textup{i}\infty}\sum_{\nu=0}^{1}
\Big(  (-1)^{C} \,  S_{1+\nu_3 + \nu_4 + \nu}(-y-\alpha_{12s}) S_{1+ \nu }(-y-\alpha_{s34})
\\ \nonumber && \makebox[1em]{} \times
S_{1+\nu_2+\nu_3+a_s+a_t+ \nu }(-y-\alpha_{23t})
 S_{1+\nu_2+\nu_4+ \nu }(-y-\alpha_{1t4})  \,
S_{\nu_4+ \nu }(\alpha_{1234} +y)
\\ && \makebox[1em]{}  \times
S_{\nu_2+\nu_3+\nu_4 + \nu+ a_2+a_t}(\alpha_{st13} +y)
S_{\nu_2 + \nu + a_2 }(\alpha_{st24} +y) \,  S_{\nu_3 + \nu + a_s}(2Q +y) \Big){dy\over 2\textup{i}},
\label{RW_s_a2qq}\end{eqnarray}
where
\bea\label{csign}
(-1)^{C}=(-1)^{\nu\left(a_sa_t+a_2a_t+a_s\nu_2+a_2\nu_3+a_2\nu_4+a_t\nu_4+a_2+a_t+a_s\right)}\, ,
\eea
and
\bea\nonumber &&
B=a_sa_t(\nu_3+\nu_2)+a_s(\nu_2\nu_3+\nu_2+\nu_3)+a_t(\nu_2\nu_4+\nu_3\nu_4+\nu_2+\nu_3+\nu_4)
\\ \nonumber && \makebox[1em]{}
+a_2a_t(\nu_2+\nu_3+\nu_4+1)+a_2(\nu_2\nu_3+\nu_2\nu_4+\nu_3\nu_4+\nu_2+\nu_3+\nu_4+1)
\\ \nonumber && \makebox[1em]{}
+\nu_2\nu_3\nu_4+\nu_2\nu_3+\nu_2\nu_4+\nu_3\nu_4+\nu_2+\nu_4\, .
\eea

Taking $\nu_1$ as in (\ref{nu1u}) and  expressing $a_1, a_4$ in terms of $a_2$, $a_s$ and $a_t$ with the
help of relations (\ref{aaa1}), we find that the sign $(-1)^C$ (\ref{csign}) coincides 
with $(-1)^X$ (\ref{signX}).
So, identifying $y=-u$, we see that function (\ref{generlam}) coincides with (\ref{RWdlo}) up to the
indicated $\mathcal{P}(\alpha_j, \nu_j)$ factor and $(-1)^B$ sign. Note that the sign $(-1)^X$
was found in \cite{Pawelkiewicz:2013wga}  only by analyzing a degeneration limit of (\ref{RW_s_a2}).
Here it is obtained from the condition that the Racah-Wigner symbols are written as the parafermionic
hypergeometric integral (\ref{jparker}) for $r=2$. An important fact is that now we have
in formula (\ref{generlam}) six independent discrete variables, say, $\nu_j, a_j,\, j=1,2,3,$
as it should be for the parafermionic hypergeometric function.

We can apply also relation (\ref{fertv}) in the opposite direction to the integral \eqref{generlam} and derive
the generalization of integral \eqref{rwhps}, giving, up to some prefactors, the Racah-Wigner symbols in the  Ponsot-Teschner type parametrization \eqref{ptparam},
for  non-zero $a_i$:
\be
\tilde J_0(\underline{\alpha},\underline{\nu},\underline{a})
=(-1)^{\nu_2+\nu_4+a_2-a_s}\Omega^{-1}(\underline{\alpha},\underline{\nu},\underline{a})
I_{\alpha_s^{a_s},\alpha_t^{a_t}}\left[\begin{array}{cc}
\alpha_3^{a_3}& \alpha_2^{a_2}\\
\alpha_4^{a_4}& \alpha_1^{a_1} \end{array}\right]_{\nu_1\nu_2}^{\nu_3\nu_4},
\ee
where
\bea \nonumber &&
I_{\alpha_s^{a_s},\alpha_t^{a_t}}\left[\begin{array}{cc}
\alpha_3^{a_3}& \alpha_2^{a_2}\\
\alpha_4^{a_4}& \alpha_1^{a_1} \end{array}\right]_{\nu_1\nu_2}^{\nu_3\nu_4}=\int_{-\textup{i}\infty}^{\textup{i}\infty}\sum_{\nu=0}^{1}
\Lambda(y+\gamma_1^{\circ},\nu_3+a_2+a_s+a_t+\nu)
\\ \nonumber && \makebox[1em]{}  \times
\Lambda(y+\gamma_2^{\circ},\nu+\nu_4+a_2)\Lambda(y+\gamma_3^{\circ},\nu+\nu_3+a_s)\Lambda(y+\gamma_4^{\circ},\nu-\nu_4+a_s)
\\ \nonumber &&  \makebox[1em]{} \times
\Lambda(-y+\beta_1^{\circ},-\nu-1-a_s)
\Lambda(-y+\beta_2^{\circ}, -\nu+1+\nu_2-\nu_3-a_2-a_s-a_t)
\\  &&  \makebox[1em]{} \times
\Lambda(-y+\beta_3^{\circ},-\nu+1-\nu_2-\nu_3-a_2-a_s)
\Lambda(-y+\beta_4^{\circ},-\nu-1)\frac{dy}{2\textup{i}}.
\label{jrww} \eea
Using formulae in Appendix B, we can rewrite this expression in terms of the $S_K(x)$-functions:
\bea\nonumber &&
I_{\alpha_s^{a_s},\alpha_t^{a_t}}\left[\begin{array}{cc}
\alpha_3^{a_3}& \alpha_2^{a_2}\\
\alpha_4^{a_4}& \alpha_1^{a_1} \end{array}\right]_{\nu_1\nu_2}^{\nu_3\nu_4}=(-1)^{F}\int_{-\textup{i}\infty}^{\textup{i}\infty}\sum_{\nu=0}^{1} (-1)^{E}
S_{1+\nu+\nu_3+a_4}(y+\gamma_1^{\circ})
\\ \nonumber &&  \makebox[1em]{} \times
S_{1+\nu+\nu_4+a_2}(y+\gamma_2^{\circ})S_{1+\nu+\nu_3+a_s}(y+\gamma_3^{\circ})S_{1+\nu+\nu_4+a_s}(y+\gamma_4^{\circ})
S_{\nu-a_s}(-y+\beta_1^{\circ})
\\  && \makebox[1em]{} \times
S_{\nu+\nu_2+\nu_3-a_4}(-y+\beta_2^{\circ})S_{\nu+\nu_2+\nu_3-a_1}(-y+\beta_3^{\circ}) S_{\nu}(-y+\beta_4^{\circ})\frac{dy}{2\textup{i}},
\label{newform}\eea
where
$ E=\nu(\nu_2+\nu_4)+\nu(\nu_3 a_2+\nu_2a_t+\nu_4a_2+\nu_4a_s+a_2a_s+a_s+a_t) $
and
\bea\nonumber &&
F=\nu_3\nu_2+\nu_4+\nu_2\nu_3a_t+\nu_2a_2a_t+\nu_2a_sa_t+\nu_3a_2a_s+\nu_3a_t+
\\ && \makebox[4em]{}
\nu_2a_s+\nu_2a_2+\nu_4a_s+a_2a_t+a_sa_t+a_s\, ,
\nonumber \\ \nonumber  &&
\Omega(\underline{\alpha},\underline{\nu},\underline{a})=(-1)^{T}
S_{\nu_3}(Q+\alpha_s-\alpha_3-\alpha_4)S_{\nu_4}(Q+\alpha_s-\alpha_1-\alpha_2)
\\ \nonumber && \makebox[2em]{} \times
S_{\nu_2+a_3}(Q-\alpha_t+\alpha_2-\alpha_3)
S_{\nu_1+a_1}(Q-\alpha_t+\alpha_4-\alpha_1)
\\ \nonumber && \makebox[2em]{}\times
S_{\nu_2+a_t}(-Q+\alpha_t+\alpha_2+\alpha_3)S_{\nu_3+a_4}(Q-\alpha_s+\alpha_3-\alpha_4)
\\ &&\makebox[2em]{}\times S_{\nu_1+a_t}(-Q+\alpha_t+\alpha_4+\alpha_1)S_{\nu_4+a_2}(Q-\alpha_s+\alpha_1-\alpha_2),
\label{omega2}\eea
with the integer $T$ of the form
\bea &&
T=\nu_3(\nu_2+\nu_4)+(a_s+a_t+a_2)(\nu_2\nu_3+\nu_2\nu_4+\nu_4\nu_3)+a_s(\nu_2+1)
\\ \nonumber && \makebox[2em]{}
+a_t(\nu_4+1)
+a_2a_s(\nu_3+1)+a_2a_t(\nu_4+1)+a_ta_s(\nu_3+1)+a_ta_sa_2\, .
\eea

To conclude, the parafermionic hypergeometric function \eqref{jparker} for $r=2$ coincides
with the key functional part of the universal Racah-Wigner symbols in $N=1$
supersymmetric Liouville theory determined in \cite{Pawelkiewicz:2013wga}.
It is natural to expect that for arbitrary $r$ it will determine key functional ingredient of the
fusion matrices of the general parafermionic Liouville field theory of \cite {Bershtein:2010wz}
and corresponding generalization of the Racah-Wigner symbols.

\vspace{0.5cm}

{\bf\large Acknowledgements.}
This study has been partially funded within the framework of the HSE University Basic Research Program.
The work of Elena Apresyan was supported by  Armenian SCS grants 21AG‐1C024 and 20TTAT-QTa009.

\appendix

\section{Sign calculations I}

Here we compute the signs appearing in passing from the $\Lambda(x,K)$-functions to $S_{K+1}(x)$-functions
in the derivation of the integrals \eqref{inttv85} and \eqref{RW_s_a2qq}. After cumbersome calculations we obtain:
\bea\nonumber
&&\Lambda(-y-\alpha_{23t}, \nu_2-\nu_3-\nu-a_s-a_t)=S_{\nu_2-\nu_3-\nu-a_s-a_t+1}(-y-\alpha_{23t})
\\ \nonumber
&&\makebox[3em]{} \times
(-1)^{a_sa_t(\nu_3+\nu+\nu_2)+(a_s+a_t)
(\nu_2\nu+\nu_3\nu+\nu_2\nu_3+\nu_2+1)+\nu_2\nu_3\nu+\nu_2\nu+\nu+\nu_2\nu_3+\nu_3}\, ,
\\  && \nonumber
\Lambda(-y-\alpha_{14t}, \nu_2-\nu_4-\nu)=S_{\nu_2-\nu_4-\nu+1}(-y-\alpha_{14t})
(-1)^{\nu_2\nu_4\nu+(\nu_2+1)(\nu+\nu_4)},
\\ \nonumber &&
\Lambda(-y-\alpha_{12s}, -\nu_3-\nu_4-\nu)=S_{-\nu_3-\nu_4-\nu+1}(-y-\alpha_{12s})(-1)^{(\nu_3+\nu_4+\nu)(\nu_3\nu_4\nu+1)}\, ,
\eea
\bea\nonumber && 
\Lambda(-y-\alpha_{34s},-\nu)=S_{\nu+1}(-y-\alpha_{34s})(-1)^\nu\, ,
\\ \nonumber  &&  
\Lambda(y+\alpha_{1234}, \nu_4+\nu+1)=S_{\nu_4+\nu}(-1)^{\nu_4\nu}\, ,
\eea
\bea \nonumber &&
\Lambda(y+\alpha_{13st}, -\nu_2+\nu_3+\nu_4+\nu-1+a_2+a_t)=
S_{-\nu_2+\nu_3+\nu_4+\nu-1+a_2+a_t}(y+\alpha_{13st})\\ \nonumber
&& \makebox[3em]{}
=(-1)^{\nu[a_2a_t+(a_2+a_t)
(\nu_2+\nu_3+\nu_4+1)+\nu_2\nu_4+\nu_3\nu_4+\nu_2\nu_3+\nu_3+\nu_4+1]+A},
\\ \nonumber &&
A=a_2a_t(\nu_2+\nu_3+\nu_4+1)+(a_2+a_t)(\nu_2\nu_3+\nu_2\nu_4+\nu_3\nu_4+\nu_3+\nu_4+1)
\\ \nonumber && \makebox[3em]{}
+\nu_2\nu_3\nu_4+\nu_3\nu_4+\nu_2+\nu_3+\nu_4+1\, ,
\\  \nonumber &&   \makebox[4em]{}
\Lambda(y+2Q, \nu_3+\nu-1+a_s)=S_{\nu_3+\nu+a_s}(y+2Q)(-1)^{(\nu_3+1)(\nu+1)( a_s+1)}\, ,
\\ \nonumber  &&  \makebox[4em]{}
\Lambda(y+\alpha_{24st}, -\nu_2+\nu+1-a_2)=S_{-\nu_2+\nu-a_2}(y+\alpha_{24st})(-1)^{(\nu+1)\nu_2a_2}\, .
\eea

\section{Sign calculations II}

Here we compute the signs that emerge after the replacements of $\Lambda(x,K)$-functions
by $S_{K+1}(x)$ during the derivation of the expression \eqref{newform}.
After lengthy calculations we obtain:
\bea \nonumber
&&\Lambda(y+\gamma_1^{\circ},\nu_3+a_2+a_s+a_t+\nu)
=S_{\nu_3+a_2+a_s+a_t+\nu+1}(y+\gamma_1^{\circ})
\\ \nonumber && \makebox[3em]{} \times
(-1)^{\nu\nu_3(a_t+a_s+a_2)+(\nu+\nu_3)(a_2a_t+a_sa_t+a_2a_s)+a_2a_sa_t},
\\ \nonumber &&
\Lambda(y+\gamma_2^{\circ},\nu+\nu_4+a_2)=S_{\nu+\nu_4+a_2+1}(y+\gamma_2^{\circ})(-1)^{\nu\nu_4a_2},
\\ \nonumber &&
\Lambda(y+\gamma_3^{\circ},\nu+\nu_3+a_s)=S_{\nu+\nu_3+a_s+1}(y+\gamma_3^{\circ})(-1)^{\nu\nu_3a_s},
\eea
\bea\nonumber &&
\Lambda(y+\gamma_4^{\circ},\nu-\nu_4+a_s)=S_{\nu-\nu_4+a_s+1}(y+\gamma_4^{\circ})(-1)^{\nu_4(\nu+1)(a_s+1)},
\\ \nonumber &&
\Lambda(-y+\beta_1^{\circ},-\nu-1-a_s)=S_{-\nu-a_s}(-y+\beta_1^{\circ})(-1)^{(\nu+1)(a_s+1)},
\eea
\bea \nonumber &&
\Lambda(-y+\beta_2^{\circ}, -\nu+1+\nu_2-\nu_3-a_2-a_s-a_t)
=S_{ -\nu+\nu_2-\nu_3-a_2-a_s-a_t}(-y+\beta_2^{\circ})
\\ \nonumber && \makebox[2em]{} \times
(-1)^{\nu\nu_3(a_t+a_s+a_2+1)+(\nu+\nu_3)(a_2a_t+a_sa_t+a_2a_s+a_2+a_s+a_t)+a_2a_t+a_sa_t+a_2a_s+a_2a_sa_t}
\\ \nonumber && \makebox[2em]{} \times
(-1)^{\nu_2[(\nu+\nu_3)(a_2+a_s+a_t)+\nu\nu_3+a_2a_t+a_sa_t+a_2a_s]},
\\ \nonumber &&
\Lambda(-y+\beta_3^{\circ},-\nu+1-\nu_2-\nu_3-a_2-a_s)=S_{-\nu-\nu_2-\nu_3-a_2-a_s}(-y+\beta_3^{\circ})\\ \nonumber
&& \makebox[1em]{}  \times
(-1)^{\nu\nu_3(a_2+a_s+1)+(\nu+\nu_3)(a_2+a_s+a_2a_s)+a_2a_s}
(-1)^{\nu_2[\nu\nu_3+(\nu+\nu_3)(a_2+a_s+1)+a_2a_s+a_2+a_s]},
\\ \nonumber  && \makebox[6em]{}
\Lambda(-y+\beta_4^{\circ},-\nu-1)=(-1)^{\nu+1}S_{-\nu}(-y+\beta_4^{\circ}).
\eea

\end{document}